\documentclass[aps,pra,reprint,superscriptaddress]{revtex4-1}

\usepackage[utf8]{inputenc}
\usepackage[english]{babel}
\usepackage{physics}
\usepackage{graphicx}
\usepackage{mathtools}
\usepackage{bm}
\usepackage{epstopdf}
\usepackage{siunitx}
\usepackage{amsfonts}
\usepackage{amssymb}
\usepackage{color,}

\makeatletter
\renewcommand*{\@fnsymbol}[1]{\ensuremath{\ifcase#1\or \dagger\or *\or * \else\@ctrerr\fi}}
\makeatother

\begin{document}

\title{Realization of efficient quantum gates with a superconducting qubit-qutrit circuit}

\author{T. Bækkegaard}
\author{L.B. Kristensen}
\author{N.J.S. Loft}
\affiliation{Department of Physics and Astronomy, Aarhus University, DK-8000 Aarhus C, Denmark.}
\author{C.K. Andersen}
\affiliation{Department of Physics, ETH Zürich, CH-8093 Zürich, Switzerland}
\author{D. Petrosyan}
\affiliation{Institute of Electronic Structure and Laser, FORTH, GR-71110 Heraklion, Greece.}
\author{N.T. Zinner}
\email[correspondance should be addressed to N.T.Z. at ]{zinner@phys.au.dk}
\affiliation{Department of Physics and Astronomy, Aarhus University, DK-8000 Aarhus C, Denmark.}
\affiliation{Aarhus Institute of Advanced Studies, Aarhus University, DK-8000 Aarhus C, Denmark.}

\date{\today}

\begin{abstract}
\noindent Building a quantum computer is a daunting challenge since it requires good control but also good isolation from the environment to minimize decoherence. It is therefore important to realize quantum gates efficiently, using as few operations as possible, to reduce the amount of required control and operation time and thus improve the quantum state coherence.
Here we propose a superconducting circuit for implementing a tunable system consisting of a qutrit coupled to two qubits.
This system can efficiently accomplish various quantum information tasks, including generation of entanglement of the two qubits and conditional three-qubit quantum gates, such as the Toffoli and Fredkin gates. Furthermore, the system realizes a conditional geometric gate which may be used for holonomic (non-adiabatic) quantum computing. The efficiency, robustness and universality of the presented circuit makes it a promising candidate to serve as a building block for larger networks capable of performing involved quantum computational tasks.
\end{abstract}


\maketitle

\mbox{ }
\clearpage



\noindent Richard Feynman famously suggested to simulate quantum physics
with quantum computers \cite{Feynman1982}. Fourteen years later, Seth Lloyd
proved that an array of spins with tunable interactions indeed represents
a universal quantum simulator \cite{Lloyd1073}.
Dynamically controlled spin chains can realize analog quantum simulations and digital quantum computations.
Several physical systems are being explored for implementing tunable spin chains
in the quantum regime, including trapped ions and atoms \cite{PhysRevLett.92.207901, Blatt2012, Khajetoorians1062}, quantum dots \cite{RevModPhys.85.961} and superconducting circuits \cite{PhysRevA.75.032329,PhysRevLett.89.117901}. Over the last decade, superconducting circuits have steadily improved to become one of the most prominent candidates for the realization of scalable quantum computing \cite{Blais_cQED,Devoret1169_QIP_outlook,Wendin_2017_review,GU20171}. With the development of the transmon qubit \cite{transmon_original} and further advances, such as the 3D transmon \cite{Devoret1169_QIP_outlook}, coherence times above $44\, \si{\micro\second} $ \cite{PhysRevLett.111.080502,PhysRevB.86.100506,Kandala2017,Wang2018_cavity_attenuators} and per-step multi-qubit gate fidelity at the fault tolerance threshold for surface code error correction \cite{PhysRevA.86.032324} have been achieved on multi-qubit devices \cite{surface_code_threshold}.

Many approaches for entangling quantum gates with superconducting qubits have been implemented experimentally \cite{Yamamoto2003,DiCarlo2009,PhysRevLett.113.220502,PhysRevApplied.6.064007,PhysRevLett.119.180511,PhysRevA.93.060302,Reagoreaao3603,surface_code_threshold,Neeley2010} and many more have been proposed theoretically \cite{PhysRevA.77.062339,PhysRevLett.108.120501,PhysRevA.94.012328,PhysRevLett.116.180501,PhysRevA.91.023828,Royer2017fasthighfidelity}.
Still, as the search for better coherence, lower error rates and faster quantum gate operation times continues, more efficient universal realizations of key operations for a quantum processor are needed. Most implementations so far have used only one- and two-qubit quantum operations for realizing important multi-qubit gates \cite{nielsen_chuang_2010} such as the three-qubit quantum Toffoli \cite{Toffoli1980} (\textsc{ccnot}) and Fredkin \cite{Fredkin1982} (\textsc{cswap}) gates, requiring a theoretical minimum of five two-qubit gates \cite{physRevA.91.032302_fredkin_minimum, PhysRevA.88.010304_toffoli_minimum, nielsen_chuang_2010}. This large number of required gates can be remedied by the use of a higher-lying state of a qutrit which can simplify the implementation of e.g. the Toffoli gate to three two-qubit gates as implemented optically in \cite{Lanyon2008} and in superconducting circuits in \cite{implementation_toffoli}. Moreover, the current implementations are highly specialized, meaning that the fabricated superconducting circuit is used just to implement a single three-qubit gate. A single circuit implementing several important universal quantum gates with high fidelity and minimal external control is therefore desirable.

Here, we propose a superconducting circuit realizing two qubits and a qutrit in between. First we show how the circuit can be used to generate two- and three-qubit entangled states \cite{entanglement_role, nielsen_chuang_2010, DiCarlo2010_three_qubit_entanglement}. Then we discuss how to implement the Fredkin gate using only two three-qubit operations, and the Toffoli gate using two (one-qubit) Hadamard gates and a three-qubit gate employing the intrinsic Ising-like $ZZ$ (longitudinal) couplings and an external one-qubit driving. Finally we discuss how to implement a double-controlled universal unitary single-qubit gate. To illustrate this capability, we use the $ZZ$-couplings to realize the double-controlled holonomic gate \cite{Duan1695,HQC}. The geometric nature of the holonomic gate provides robustness but, the first proposals required adiabatic control leading to more time for errors to occur. The increase in errors was partly avoided by the introduction of decoherence-free subspaces \cite{PhysRevLett.95.130501}, which can significantly reduce the detrimental effects of noise. We will instead implement a non-adiabatic generalization \cite{Sjoqvist_NHQC}, circumventing many of these difficulties. Furthermore, we will show that this double-controlled gate can be used to implement the three-qubit Deutsch gate and is therefore universal for quantum computing in itself, adding yet another tool in our toolbox for efficient quantum gates.


\section* {Results} \label{sec:results}
\subsection*{Effective Hamiltonian of the system} \label{subsec:formalism}

\begin{figure*}[htb!]
  \centering
  \includegraphics[width=\textwidth]{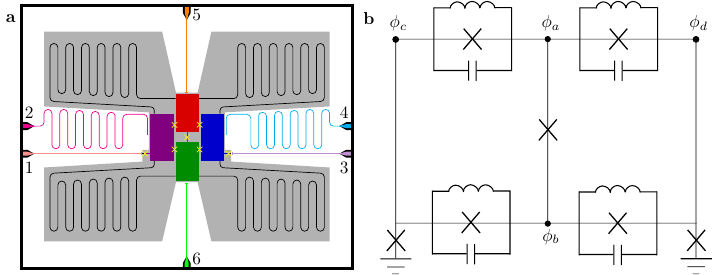}
  \caption{
    \textbf{a} Sketch of a possible physical implementation of the proposed circuit. Each colored box is a superconducting island corresponding to a node in a lumped circuit element model. Josephson junctions are shown schematically as yellow crosses. Bent black wires are inductors. The numbered colored lines are controls for readout and driving of the circuit: 1 and 3 are the flux lines for frequency tuning of the outer qubits, 2 and 4 are resonators capacitively coupled to left and right qubits, and lines 5 and 6 are control and driving of the two middle islands forming the middle qutrit.
    \textbf{b} Effective Lumped circuit scheme of the same circuit. The four nodes in the system are shown as dots and Josephson Junctions are shown as crosses.}
\label{fig:circuit}
\end{figure*}

\noindent Consider the circuit with four connected superconducting islands and the corresponding effective lumped circuit element model in Figure \ref{fig:circuit}.
The top-bottom symmetry between the capacitors and the Josephson junctions cancels the direct exchange interaction mediated by the capacitances and the leading order term from the Josephson junctions. On the other hand, we choose an asymmetry in the inductive couplings such that the difference between them controls the Heisenberg-like exchange coupling, which can be made arbitrarily small if so desired. 
Furthermore, while the exchange interaction mediated by the leading-order Josephson term ($E_{J1}/2$ in Supplemental Note 1) is canceled by symmetry, the dispersive ($ZZ$) coupling survives. The couplings are defined in Eq. (A61) and as seen in Supplementary Note 2, a $ZZ$ coupling strength of $\sim \SI{30}{\mega\hertz} $ is realistic. Moreover, the $ZZ$ coupling can be tuned in-situ by an external flux.

Using the effective lumped-element circuit of Figure \ref{fig:circuit}b and following the standard procedure for circuit quantization \cite{Devoret,transmon_original}, we derive the Hamiltonian of the system involving a suitable set of variable for the relevant dipole modes of the circuit, as detailed in the Method and in more details in the Supplementary Note 1.
The Hamiltonian of the resulting qubit-qutrit-qubit system shown in Figure 2 takes the form:
\begin{equation}
  \label{eq:H_full}
  \begin{split}
    H = & \frac{1}{2} \Delta_L \sigma^z_L + \Delta_M \ket{1} \bra{1}
    + \left( \Delta_M + \delta_M \right) \ketbra{2}
    + \frac{1}{2} \Delta_R \sigma^z_R \\
    & + J_{LM_{01}}\left( \sigma_L^- \ketbra{1}{0}
    + \sigma_L^+ \ketbra{0}{1} \right) \\
    & + J_{RM_{01}}\left( \sigma_R^-\ketbra{1}{0}
    + \sigma_R^+\ketbra{0}{1} \right) \\
    & + J_{LM_{12}} \left( \sigma_L^-\ketbra{2}{1}
    + \sigma_L^+\ketbra{1}{2} \right) \\
    & + J_{RM_{12}}\left( \sigma_R^-\ketbra{2}{1}
    + \sigma_R^+\ketbra{1}{2} \right) \\
    & + J_{LM}^{(z)} \sigma^z_L\left( D_1 \ketbra{1} + D_2 \ketbra{2} \right) \\
    & + J_{RM}^{(z)}\sigma^z_R \left( D_1\ketbra{1} + D_2\ketbra{2} \right) ,
  \end{split}
\end{equation}
where $\sigma_{\alpha}^+$ and  $\sigma_{\alpha}^-$ are the spin-1/2 raising
and lowering operators for the left ($\alpha =L$) and right ($\alpha =R$)
qubits, $\sigma_{\alpha}^z $ is the Pauli $Z$ operator,
and $\Delta_{L,R}$ is the energy differences between the spin-up and spin-down
states of the corresponding qubit. The states of the qutrit are denoted by
$\ket{j}$ ($j =0,1,2$), $\Delta_{M}$ is the energy of state $\ket{1}$ and
$\Delta_{M} + \delta_{M}$ is the energy of state $\ket{2}$, making the anharmonicity equal to $ \Delta_M - \delta_M $, with the energy of the ground state $\ket{0}$ set to zero. $\Delta_{M}$ and $\delta_{M}$ can be tuned dynamically by an external flux if additional flux lines are added to the circuit, or by an AC-Stark shift  stemming from off-resonant microwave driving \cite{Blais_cQED} using the lines 5 and 6 in fig. \ref{fig:circuit} (a) (see Supplementary Note 1).
We note that the AC-Stark shift is included here for additional dynamical tuning of the circuit but it is not essential for the gate implementations discussed below. 

\begin{figure}[htb!]
  \centering
  \includegraphics[width=0.99\columnwidth]{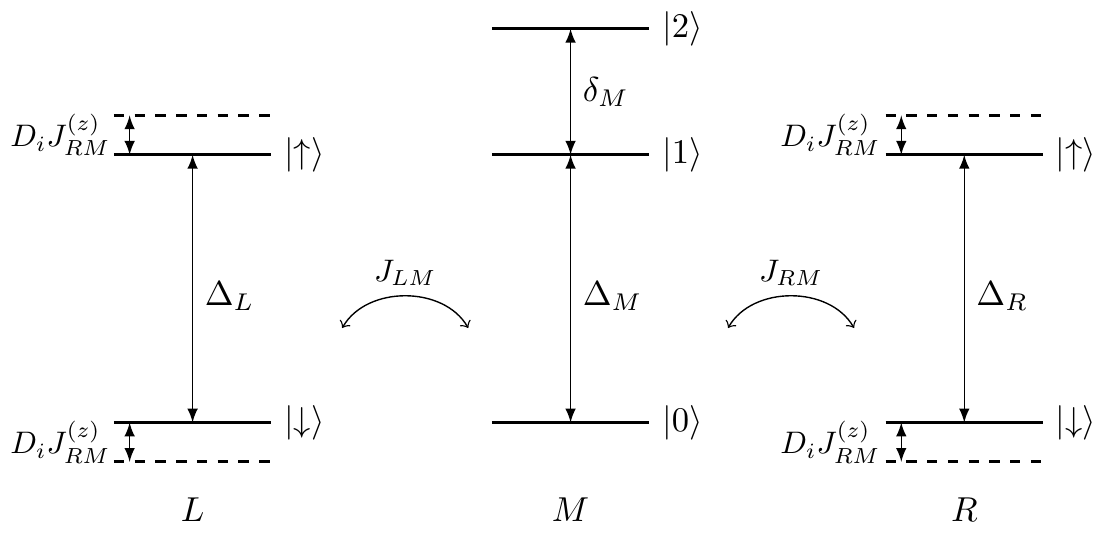}
  \caption{
    Energy diagram of the system of two qubits (left, $L$, and right, $R$) and a qutrit (middle, $M$) described by the Hamiltonian in Equation \eqref{eq:H_full}. Also shown are the exchange couplings $J_{\alpha M}$ and state-dependent energy shifts $J_{\alpha M}^{(z)}$ of \eqref{eq:H_full}. $D_i$ depend on the state of the qutrit $\ket{i}$, with, $D_0=0$, and typically, $D_{1} \gtrsim 2$ and $D_{2} \lesssim 4$.}
\label{fig:diagram}
\end{figure}

The exchange ($XY$) coupling strengths between the qubits and the qutrit are given by
$J_{\alpha M}$. Typically, the coupling $J_{\alpha M_{12}}$ ($ \alpha = L,R$) to the
$\ket{1} \leftrightarrow \ket{2}$ transition is stronger than the coupling $J_{\alpha M_{01}}$ to the $\ket{0} \leftrightarrow \ket{1}$ transition by a factor $\sim \sqrt{2}$. The coefficients $J_{\alpha M}^{(z)}$ determine the dispersive ($ZZ$) interaction between the qubits and the qutrit with  $D_1 \gtrsim 2$ and $D_2 \lesssim 4$, with the two parameters converging at 3 as the mixing is increased via the off-resonant driving. For clarity, these parameters are shown in Figure \ref{fig:diagram}.
In a perfectly left-right symmetric circuit, we have $\Delta_L = \Delta_R$ and $J_{L M} = J_{R M}$ for both qutrit transitions which is assumed below unless otherwise stated.

For typical experimental parameters, the coupling strengths $J$'s are in the range
of few to some tenths of MHz, while the energies $\Delta$'s and
$\delta$'s are in the $~10$ GHz range. We choose realistic values of the experimental parameters so as to stay within the transmon regime \cite{transmon_original}.


Apart from the intrinsic dynamics of the system, we will employ an external microwave (mw) field of (variable) frequency $\omega_{\mathrm{mw}}$ to drive the system. Physically, the driving can be applied through resonators 2 and 4 capacitively coupled to the outer qubits, and control lines 5 and 6 capacitively coupled to the qutrit, as shown in Figure 1(a). The microwave field induces transitions between the qubit and qutrit states as described by the Hamiltonian
\begin{equation} \label{eq:mw_driving}
  \begin{split}
    H_{\mathrm{mw}} =& \cos(\omega_{\mathrm{mw}}t)
      (\Omega_L \sigma^+_L + \Omega_R \sigma^+_R \\ 
    &+ \Omega_{1} \ketbra{0}{1} + \Omega_{2} \ketbra{1}{2} + \mathrm{H. c} ),
  \end{split}
\end{equation}
where $\Omega$'s are the corresponding Rabi frequencies.
Moreover, multifrequency pulses generated by an appropriate microwave source and directed to the qutrit via the control lines can be used to dynamically tune the qutrit transitions (see Supplementary Note 1 for the general treatment of transitions of a capacitively coupled qutrit).
We note that unlike our qubits and qutrit, in flux qubits optical selection rules may depend on the magnetic flux \cite{PhysRevLett.95.087001_selection_rules_flux_qubit}.

Our system can be used to achieve many quantum information tasks, examples of 
which are described below.
The qutrit can encode a qubit in either states $\left(\ket{0},\ket{1}\right)$ or states $\left(\ket{0},\ket{2}\right)$. This is solely a matter of convenience and it is straightforward to toggle between these two encodings by applying a $\pi$-pulse on the $\ket{1}\leftrightarrow\ket{2}$ transition.


\subsection*{Qutrit dissociation and entangled state preparation}
\label{subsec:dissociation}

\noindent We now discuss how to deterministically prepare entangled states in the setup, which is of great importance for quantum computation and information tasks \cite{QC_implementation,entanglement_role}. In our system, we can employ the qutrit to deterministically prepare an entangled Bell state between the outer qubits
$ \frac{1}{\sqrt{2}}(\ket{\downarrow \downarrow} + \ket{\uparrow \uparrow})$,
as detailed below.

\begin{figure}
  \includegraphics[width=0.99\columnwidth]{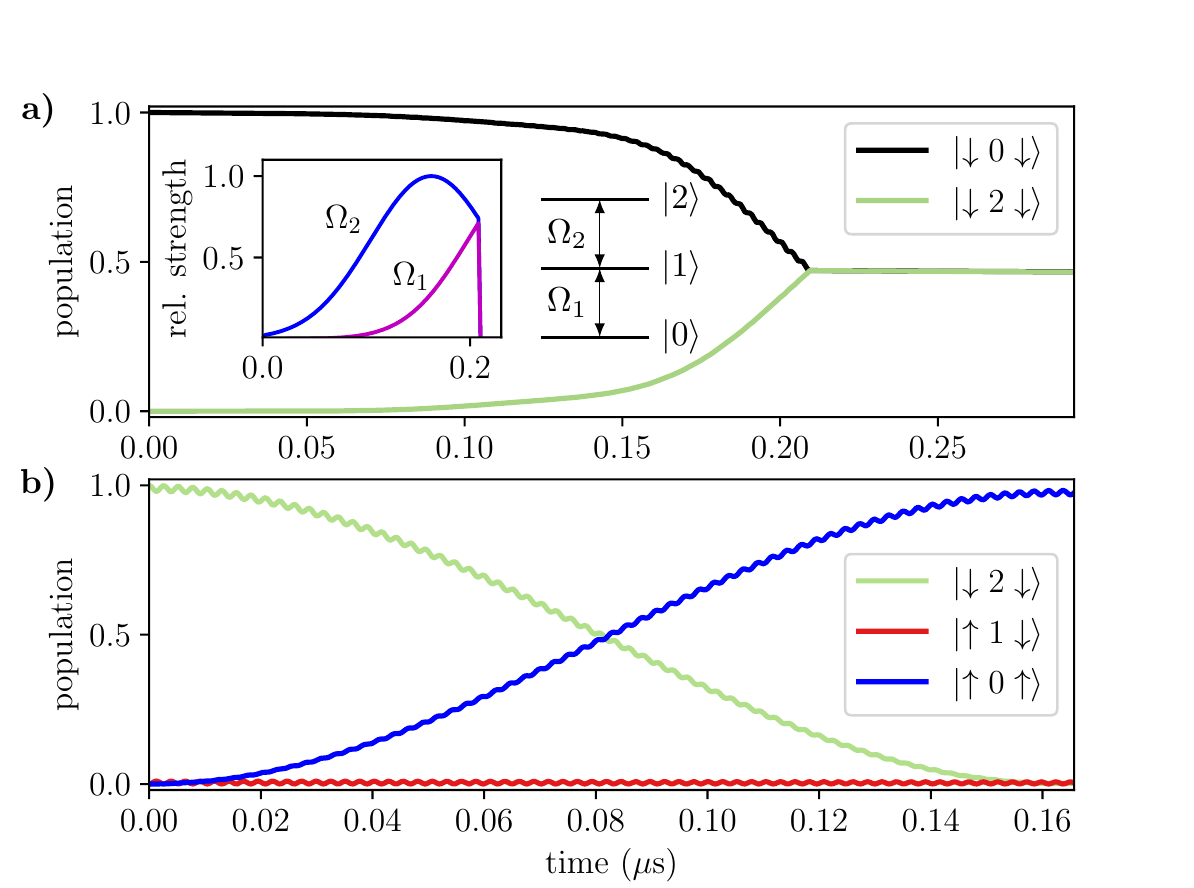}
  \caption{\label{fig:dissociation}\textbf{a)}
Populations of states $\ket{0}$, $\ket{1}$ and $\ket{2}$ of the middle qutrit
during the half STIRAP in the case of off-resonant qutrit levels and $\max(\Omega_{1,2})/2\pi = \SI{20}{\mega\hertz}$. The inset shows the envelopes of the mw pulses.
\textbf{b)} Dissociation of the initial state $\ket{\downarrow 2 \downarrow} $ into the final state $\ket{\uparrow 0 \uparrow} $ with the two-photon
resonance $\Delta_M + \delta_M = \Delta_L + \Delta_R $ while the
intermediate states $\ket{\uparrow 1 \downarrow}$, $\ket{\downarrow 1 \uparrow}$
are  off-resonant. We used the parameters $J_{\alpha M_{01}}/2\pi \simeq \SI{15.1}{\mega\hertz} $ and $J_{\alpha M_{12}}/2\pi \simeq \SI{19.4}{\mega\hertz}$. We also include finite coherence times as described in Methods.}
\end{figure}

First, we tune the energy levels of the qutrit
to make its two transitions $ \ket{0} \leftrightarrow \ket{1} $ and
$ \ket{1} \leftrightarrow \ket{2} $ non-resonant with the transitions
$ \ket{\downarrow} \leftrightarrow \ket{\uparrow}$ of the qubits,
i.e., we require that $ \abs{\Delta_M - \Delta_\alpha} \gg J_{\alpha M_{01}} $
and $ \abs{\delta_M - \Delta_\alpha} \gg J_{\alpha M_{12}} $.
Starting from the ground state $\ket{0}$, we produce the superposition state $\frac{1}{\sqrt{2}}(\ket{0} + \ket{2}) $ of the qutrit by external driving, employing the STIRAP (STImulated Raman Adiabatic Passage) sequence of pulses \cite{STIRAP}, as has been also proposed \cite{PhysRevLett.100.113601} and implemented \cite{Kumar2016} in superconducting circuits before. Namely, we drive the transitions $ \ket{0} \leftrightarrow \ket{1} $ and $ \ket{1} \leftrightarrow \ket{2} $ with resonant mw-pulses of Rabi frequencies $\Omega_1$ and $\Omega_2$. The $\Omega_2$ pulse precedes the $\Omega_1$
pulse, and we adjust the overlap between the pulses so
as to obtain the transfer to state $\ket{2}$ with minimal population of the intermediate state $\ket{1}$. The two pulses are
suddenly turned off when their amplitudes are equal, resulting in the desired superposition state. The dynamics of the qutrit under the STIRAP driving is shown in Figure \ref{fig:dissociation} (a), with the inset showing the pulse sequence.

Next, the Bell state is obtained as
$ \frac{1}{\sqrt{2}}\ket{\downarrow \downarrow} (\ket{0} + \ket{2}) \to
\frac{1}{\sqrt{2}}(\ket{\downarrow \downarrow} + \ket{\uparrow \uparrow})
\ket{0}$ via ``dissociation'' of the qutrit excitation
$\ket{2}$ into two qubit excitations $\ket{\uparrow \uparrow}$.
To this end, we set $ \Delta_M + \delta_M = \Delta_L + \Delta_R$ and choose 
$\abs{\Delta_M - \Delta_\alpha} > J_{\alpha M_{01}}$ via tuning the frequencies of the outer qubits with flux control and the qutrit with the dynamical driving. Note that this condition applies when $J_{\alpha M_{01}} = J_{\alpha M_{12}}$. 
If the exchange coefficients are different, as is normally the case, the qutrit is moved out of the two-photon resonance by unequal second order level shifts
$|J_{\alpha M_{01}}|^2/(\Delta_M - \Delta_\alpha) \neq
|J_{\alpha M_{12}}|^2/(\delta_M - \Delta_\alpha)$, which can be compensated for
by adjusting $\Delta_M$ or $\delta_M$.
Making the intermediate state $\ket{1}$ non-resonant precludes its population
but prolongs the dissociation, which results in a more pronounced effect
of the noise and relaxations.
The dissociation dynamics $ \ket{\downarrow 2 \downarrow} \rightarrow \ket{\uparrow 0 \uparrow} $ is shown in Figure \ref{fig:dissociation}.

Using the pairwise $ZZ$-interactions, 
the fully entangled three-particle $GHZ$ (Green-Horne-Zeilinger) state
$\frac{1}{\sqrt{2}}(\ket{\downarrow 0 \downarrow} + \ket{\uparrow 1 \uparrow})$ can be obtained from the prepared Bell state 
$\frac{1}{\sqrt{2}}(\ket{\downarrow 0 \downarrow} + \ket{\uparrow 0 \uparrow})$
by the external driving of the middle qutrit. To this end, we apply to the circuit a weak $\pi$ pulse $\Omega_1$ which is resonant only for the
$\ket{\uparrow 0 \uparrow} \leftrightarrow \ket{\uparrow 1 \uparrow}$
transition and non-resonant for the
$\ket{\downarrow 0 \downarrow} \leftrightarrow \ket{\downarrow 1 \downarrow}$
transition, due to the $ZZ$ interactions with the strengths
$J_{\alpha M}^{(z)} \gg \Omega_1$.

Alternatively, we can encode a qubit in the $\left(\ket{0},\ket{2}\right)$ states of the qutrit and produce a different maximally entangled state
$ \frac{1}{\sqrt{2}}(\ket{\downarrow 2 \downarrow} + \ket{\uparrow 0 \uparrow})$, equivalent to the $GHZ$ state above. Starting from the simple initial state $\ket{\downarrow 2 \downarrow}$, we use only the intrinsic system dynamics by tuning the parameters until $\Delta_L = \Delta_R = \Delta_M = \delta_M$ and
$D_2J_{\alpha M}^{(z)} = \frac{2\sqrt{6}J_{\alpha M_{01}}}{\sqrt{(n\pi/c_n)^2 - 1}}$
for $n=1,2,3,\dots$ and $c_n = \cos[-1](\frac{(-1)^{n+1}}{8}) $.
Here, $n$ is a parameter controlling at which oscillation between the states $ \ket{\downarrow 2 \downarrow} $ and $\ket{\uparrow 0 \uparrow} $ their equal superposition is obtained (lower $n$ is quicker). The disadvantage of this scheme is that it requires very precise tuning of the interactions $J_{\alpha M}^{(z)}$. In contrast, for the method above, only the frequencies have to be adjusted, which is easier using the dynamical tuning or an equivalent flux tuning.
Further details are given in Supplementary Note 3.


\subsection*{Toffoli and CCZ gates} \label{subsec:CCNOT}

\noindent The controlled-controlled \textsc{not} (\textsc{ccnot}) gate, also called the Toffoli gate, is a reversible and universal 3-bit gate for classical computation \cite{Toffoli1980}. It performs a \textsc{not} (bit-flip) operation on the target bit if the two control bits are in state `1',  and does nothing otherwise. The Toffoli gate is an important element in many quantum algorithms, such as quantum error correction \cite{PhysRevLett.81.2152} and Shor's algorithm \cite{doi:10.1137/S0097539795293172}. It has been implemented in systems ranging from trapped ions \cite{PhysRevLett.102.040501} to superconducting circuits \cite{implementation_toffoli}, including a proposal for an implementation with static control optimized with machine learning \cite{Banchi2016}.

We can implement the \textsc{ccnot} gate with the left qubit and the middle qutrit acting as controls and the right qubit being the target. The state of the right qubit is then inverted only if the left qubit is in the spin up (excited) state $\ket{\uparrow}$ and the qutrit is in the ground state $\ket{0}$. The second control qubit is encoded in the qutrit states $\ket{0}$ and $\ket{2}$. The quantum \textsc{ccnot} gate is realized by first executing a double-controlled phase (\textsc{ccz}) gate that shifts the phase of the state $\ket{\uparrow0\uparrow}$ by $\pi$ (sign change) while nothing happens if either of the outer qubits are in the spin down state or if the qutrit is in state $\ket{2}$. The \textsc{ccnot} can then be obtained by the transformation:
$\operatorname{\textsc{ccnot}} = \mathcal{H} \cdot \mbox{\textsc{ccz}} \cdot \mathcal{H}$,
where $\mathcal{H}$ is the Hadamard gate that acts on the target qubit $R$. In practice, the Hadamard gate can be obtained by a $ \pi/2 $ rotation about the $y$ axis.

\begin{figure}
\centering
  \makebox[0.5\columnwidth][c]{\includegraphics[width=0.9905\columnwidth]{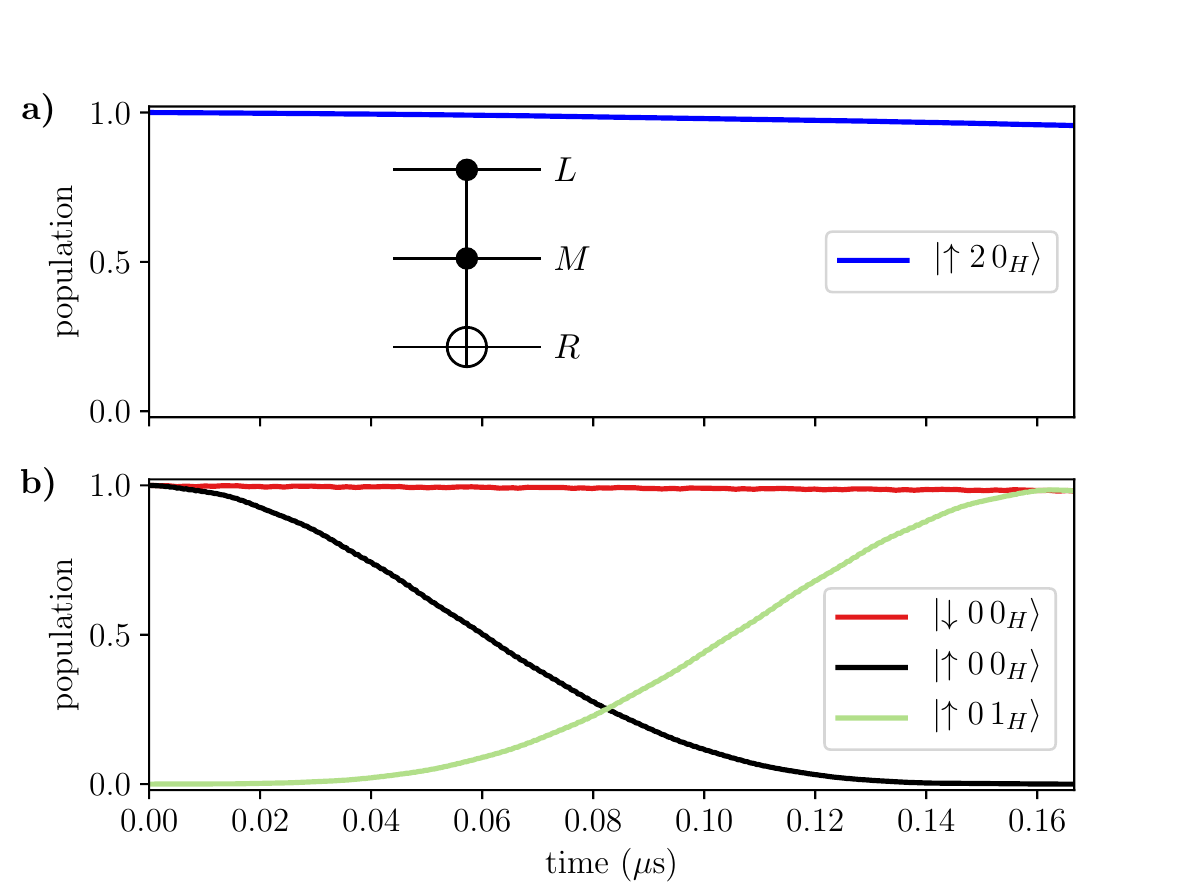}}
  \caption{Numerical simulation of the implementation of the \textsc{ccz} gate in the rotating frame. The phase of the right qubit is flipped,
$\ket{0_H} \to \ket{1_H}$, if the left qubit is in state $\ket{\uparrow}$
and the qutrit is in state $\ket{0}$, otherwise no change occurs as exemplified in a) for the state $\ket{\uparrow 2 \, 0_H} $ and in b) for the state $\ket{\downarrow 0 \, 0_H} $. A subsequent Hadamard gate on the right qubit will yield the desired \textsc{ccnot} gate.
The standard circuit representation of the Toffoli gate is shown as an inset in the upper panel of the figure.
See Supplementary Material for the parameters used in the simulation.}
  \label{fig:CCPHASE}
\end{figure}

The \textsc{ccz} gate is implemented by choosing suitable parameters such that the transitions between the qubit and qutrit states are non-resonant, while $J_{\alpha M}^{(z)}$ ($>\SI{10}{\mega\hertz}$) is large.
We apply a weak microwave field on the qutrit transition $ \ket{\uparrow 0 \uparrow} \leftrightarrow \ket{\uparrow 1 \uparrow}$ with the Rabi frequency $\Omega_1 \ll J_{\alpha M}^{(z)}$. Because of the $ZZ$ interactions, which yield a state-dependent frequency shift of the qutrit, the microwave field frequency can be chosen such that it is resonant only when both outer qubits are in the spin-up state. The microwave $2\pi$-pulse then results in the transformation
$ \ket{0} \to i \ket{1} \to - \ket{0}$ that leads to the double conditional
$\pi$ phase change of (only) the state $\ket{\uparrow 0 \uparrow}$. For simplicity, we have here used a standard square-pulse control. In a real-life implementation, a DRAG pulse \cite{PhysRevLett.103.110501_DRAG} or similar optimized pulses could be used, suppressing leakage to, and phase errors from, other levels and thus further improving the fidelity.
In Figure \ref{fig:CCPHASE} we show the results of the numerical simulations
of the \textsc{ccz} gate in the Hadamard basis for the right qubit (defined as $\ket{0_H} = \ket{+} = (\ket{\downarrow} + \ket{\uparrow})/\sqrt{2} $ and $ \ket{1_H} = \ket{-} = (\ket{\downarrow} - \ket{\uparrow})/\sqrt{2} $).
Subsequent application of the Hadamard gate to the right qubit will
complete the \textsc{ccnot} gate.
We note that because of the symmetry of the driving, we could have also chosen the qutrit state $\ket{1}$ instead of $\ket{0}$ as the `open' state, but here we can view this merely as an ancillary state.


\begin{figure*}[tb!]
\includegraphics[width=0.9\textwidth]{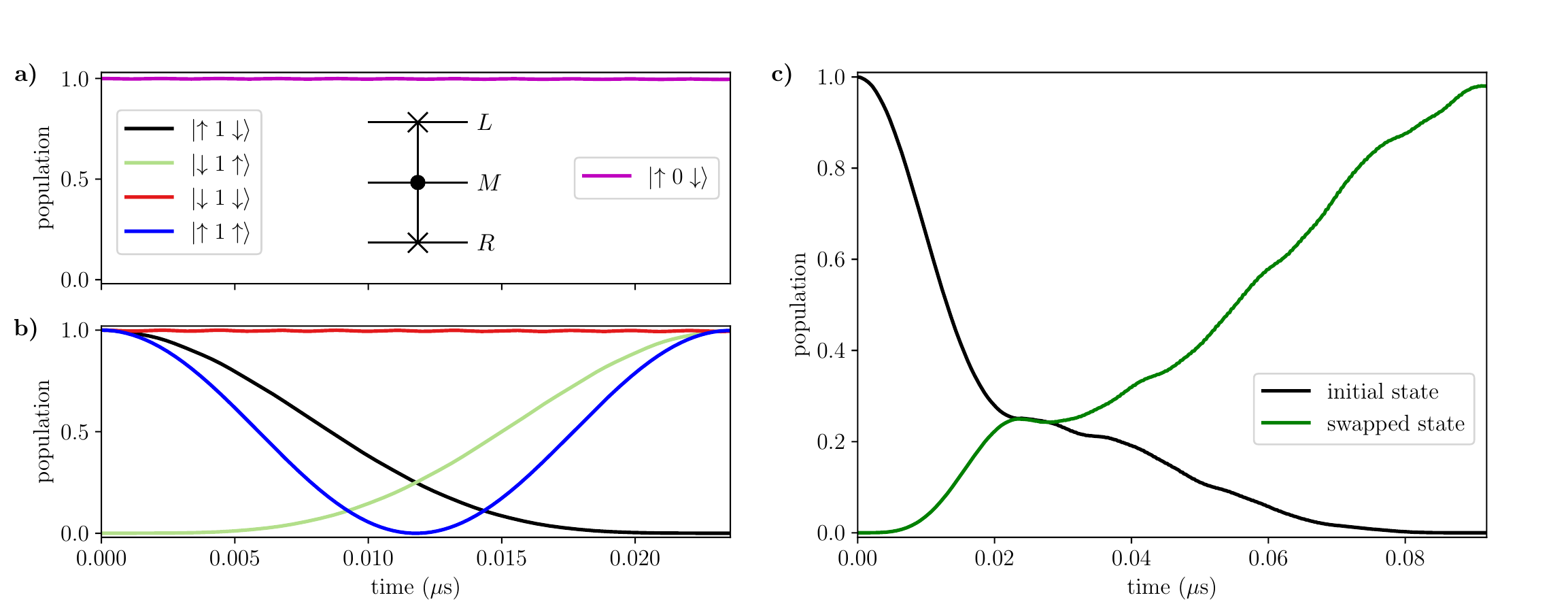}
\caption{\label{fig:CSWAP} \textbf{a)} and \textbf{b)} Numerical simulations of the \textsc{acswap} (almost \textsc{cswap})
gate for different computational basis states, with the exchange interaction $J_{\alpha M_{12}}$ resonant for time
$T = \pi / \sqrt{2}J_{\alpha M_{12}}$.
The standard circuit representation of the Fredkin gate is shown as an inset in the top part of the figure.
\textbf{b)} Numerical simulation of the full \textsc{cswap} gate
  for the initial superposition state
  $ \left[\cos(\theta_1)\ket{\uparrow} + \text{e}^{i\phi_1}\sin(\theta_1)\ket{\downarrow}\right] \ket{1} \left[\cos(\theta_2)\ket{\uparrow} + \text{e}^{i\phi_2}\sin(\theta_2)\ket{\downarrow}\right] $ with $\theta_1 = \pi/4,\ \phi_1 = 3\pi/4,\ \theta_2 = 3\pi/4 \qq{and} \phi_2 = \phi_1 $. In part 1, we perform
  the \textsc{acswap} operation during time $T_1 = \pi/2J_{\alpha M_{01}}$
  with the parameters as in figure \ref{fig:CSWAP}. In part 2 we
  perform the \textsc{ccz} gate during time $T_2 = 2\pi/ \Omega_{2}$
  with the resonant mw field of frequency
  $ \omega_\mathrm{mw} = \delta_M - 2J_{\alpha M}^{(z)}$. The full \textsc{cswap} fidelity of this state is $>0.98$, with finite coherence times included.
  See Supplementary Material for the parameters used in the simulation.}
\end{figure*}

\subsection*{Fredkin gate} \label{subsec:CSWAP}

\noindent Another classically universal 3-bit gate is the Fredkin gate, whose quantum analog
is the controlled \textsc{swap} (\textsc{cswap}) gate. Its effect is to
swap the states of the two qubits,
$\ket{\uparrow \downarrow} \leftrightarrow \ket{\downarrow \uparrow}$,
conditional upon the state of a control qubit, here encoded in the qutrit. We now use the two lowest states ($\ket{0}$ and $\ket{1}$) of the qutrit to encode the qubit such that the excited state $\ket{1}$ is `on' and the ground state $\ket{0}$ is `off'.
To realize \textsc{cswap}, we tune the energy levels of the qutrit such
that the transition $\ket{1} \leftrightarrow \ket{2}$ is resonant with
the qubit transitions $\ket{\uparrow} \leftrightarrow \ket{\downarrow}$,
i.e. $\Delta_L \simeq \delta_M \simeq \Delta_R$.
Simultaneously, the qutrit transition $\ket{0} \leftrightarrow \ket{1}$
is largely detuned, $\abs{\Delta_M - \Delta_{L,R}} \gg J_{\alpha M_{01}}$.
We then keep the resonance of the exchange interaction $J_{\alpha M_{12}} \gg J_{\alpha M}^{(z)}$ for time $T = \pi / \sqrt{2}J_{\alpha M_{12}} $. If the qutrit is in state $\ket{0}$, the qubits remain in their initial states due to absence of
resonant transitions. But if the qutrit is in state $\ket{1}$, it would
induce the swap between the qubit states,
$\ket{\uparrow 1 \downarrow} \leftrightarrow \ket{\downarrow 1 \uparrow}$,
via the resonant intermediate state
$\ket{\downarrow 2 \downarrow}$ involving the qutrit excitation.
(Resonant swap between the qubits would also occur for the
qutrit initially in state $\ket{2}$, with the intermediate
state being  $\ket{\uparrow 1  \uparrow}$.) This is illustrated
in Figure \ref{fig:CSWAP} (a).

As can also be seen in Figure \ref{fig:CSWAP} (a), however, the initial state $\ket{\downarrow 1 \downarrow}$ has trivial dynamics, unlike the rest of the swapped states
which attain a $\pi$ phase shift during the interaction time $T$. This means that we have a \textsc{swap} operation only up to a conditional phase for an arbitrary superposition input state. This is related to the phase shift of the swapped terms arising in the i\textsc{swap} gate, obtained by directly coupling two resonant qubits, which has recently attracted great interest \cite{Blais_cQED,iSWAP_IBM}. In our case with the qutrit mediating the swap,  only one state has a sign that needs correction, similarly to what Kivlichan et al. has recently called the ``fermionic simulation gate'' \cite{SWAP_google}. Because of this, we can easily mitigate this problem by using the \textsc{ccz} gate
(see Subsection ``Toffoli and CCZ gates'') to attain the $\pi$ phase shift of state $\ket{\downarrow 1 \downarrow}$ and obtain the correct \textsc{cswap} gate. In Figure \ref{fig:CSWAP} (b) we show the results of our numerical simulations of the complete
\textsc{cswap} protocol, including the conditional resonant
\textsc{swap} followed by the \textsc{ccz} gate with a total fidelity $>\SI{98}{\percent}$. More detailed
analysis is given in Supplementary Note 4.

We note that we could have equivalently performed the \textsc{cswap}
gate between the two qubits via the resonant qutrit transition
$\ket{0} \leftrightarrow \ket{1}$, while the other transition
$\ket{1} \leftrightarrow \ket{2}$ is non-resonant.
In our scheme, the qutrit has to play the role of control and thus our Fredkin gate is not a universal multi-qubit gate in itself. We could, however, imagine another qubit with controlled coupling to the qutrit as part of a larger universal circuit.


\subsection*{Double-controlled holonomic gate} \label{subsec:holonomic}

\begin{figure}
  \includegraphics[width=0.99\columnwidth]{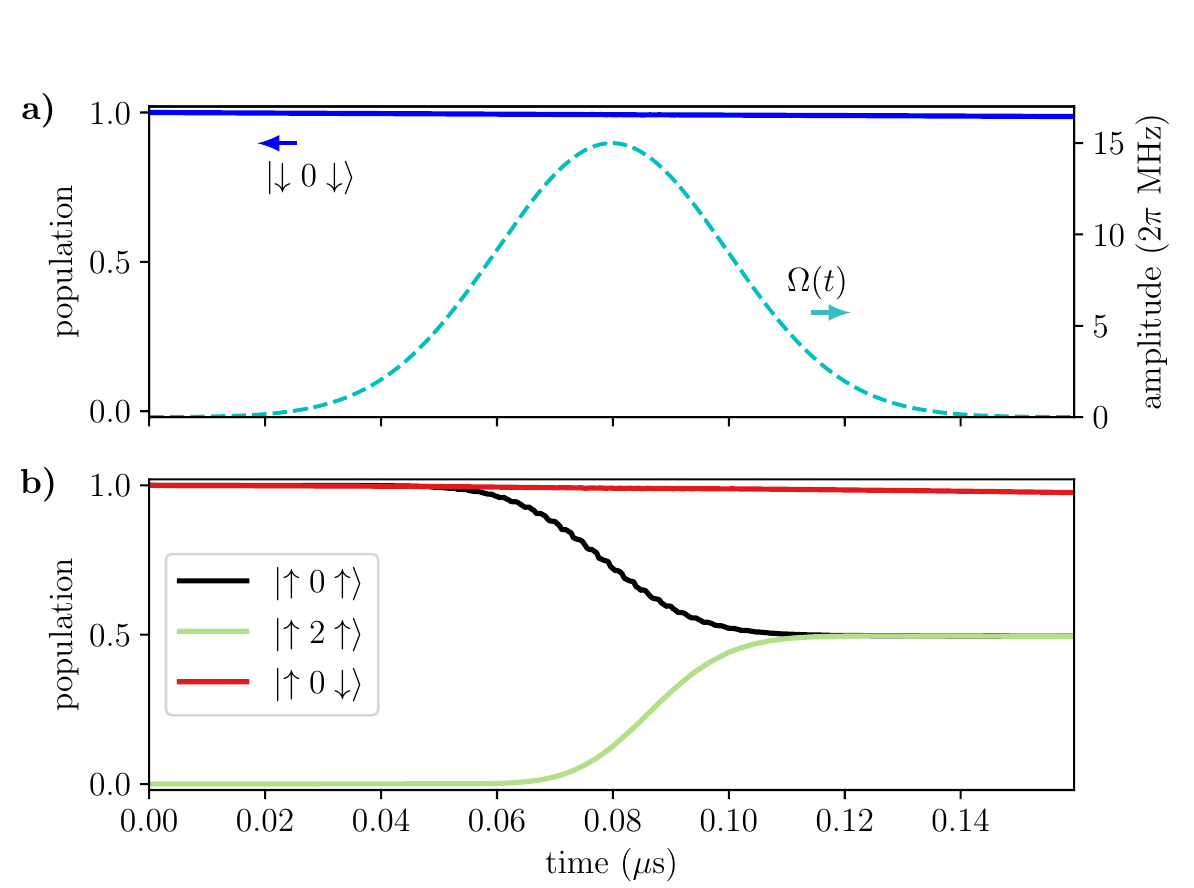}
  \caption{\textbf{a} and \textbf{b} Populations (left vertical axes) as function of time during the operation of the controlled-controlled holonomic gate in the case of $ \theta = \pi/4 $ and $ \phi = 0 $. Panel \textbf{a} also shows the envelope of the external fields plotted with dotted lines with the corresponding vertical axis on the right of the plot. See Supplementary Material for the parameters used in the simulation.}
  \label{fig:holonomic_dynamics}
\end{figure}

\noindent Another concept with importance to quantum computation \cite{nielsen_chuang_2010} is the implementation of general (non-abelian) one-qubit gates of the form (neglecting overall phase factors)
\begin{equation} \label{eq:universal_singlequbit}
  U = \begin{pmatrix}
    \text{e}^{i\phi_1}\cos(\theta) & \text{e}^{i\phi_2}\sin(\theta) \\
    -\text{e}^{-i\phi_2}\sin(\theta) & \text{e}^{-i\phi_1}\cos(\theta) 
  \end{pmatrix}.
\end{equation}
Together with a non-trivial (entangling) multi-qubit gate, they form a universal set of quantum gates \cite{nielsen_chuang_2010}.
We can implement the non-adiabatic one-qubit holonomic gate \cite{Pachos:2012:ITQ:2331123,HQC} with our qutrit, choosing states $\left(\ket{0},\ket{2}\right)$ to encode the qubit. Such gates have the advantages of being robust to parameter fluctuations due to the geometric nature, without the limitations of long gate operation times required to satisfy the adiabatic requirement \cite{PhysRevA.70.042316,doi:10.1142/S0217979201004836}. Holonomic gates have been implemented in a range of different systems \cite{holonomic_realization,PhysRevLett.110.190501}, and their stability has been well tested \cite{PhysRevLett.102.030404,PhysRevLett.112.143603}.
Choosing the same system parameters as for the \textsc{ccz} gate above, we use a driving scheme inspired by \cite{holonomic_realization} and thereby realize the single-qubit gate
\begin{equation} \label{eq:holonomic}
  U(\phi,\theta) = \begin{pmatrix}
    \cos(\theta) & \text{e}^{i\phi}\sin(\theta) \\
    \text{e}^{-i\phi}\sin(\theta) & -\cos(\theta)
  \end{pmatrix},
\end{equation}
with the computational qubit states as basis. This transformation is less general than \eqref{eq:universal_singlequbit}, but it is still universal for one-qubit rotations.

We drive the two transitions $ \ket{0} \leftrightarrow \ket{1} $ and $ \ket{1} \leftrightarrow \ket{2} $ with the external fields having the same Gaussian envelope $\Omega(t)$ but different complex coupling amplitudes $a$ and $b$, i.e. $\Omega_1(t) = a\Omega(t)$ and $\Omega_2(t) = b\Omega(t)$ in Equation \eqref{eq:mw_driving}, satisfying $\abs{a}^2 + \abs{b}^2 = 1 $. The pulse $\Omega(t)$ is turned on at time $t=0$ and turned off at
$t=\tau$, such that we get a $2\pi$-pulse, $ \int_0^\tau \Omega(t) \text{d}t = 2\pi$. Notice that this condition ensures that we end up with a closed path in parameter space and the gate is indeed holonomic.
Starting with the qutrit in the ground state $\ket{0}$, we then obtain the final
transformation $U(\phi,\theta)$ of Equation (\ref{eq:holonomic})
acting on the qutrit states $\ket{0}, \ket{2}$.
Here $\theta$ and $\phi$ are defined via
$\text{e}^{i\phi}\tan({\theta/2}) = a/b $.

By using the $J_{\alpha M}^{(z)}$ couplings to shift the qutrit frequencies, we can make the external driving field resonant or not, depending on the states of the outer qubits. This condition would then result in a controlled-controlled holonomic gate transforming the state of the qutrit according to \eqref{eq:holonomic} only when the outer (control) qubits are in e.g. the spin up state. We can show that this new gate is universal for quantum computing by first writing it in the three-qubit computational basis with $\ket{6} = \ket{\uparrow 0 \uparrow}$ and $ \ket{7} = \ket{\uparrow 2 \uparrow}$, and the rest of the 8 basis states numbered from 0 to 5:
\begin{align}
 U^\text{c}(\phi,\theta) = \left(\begin{array}{c | c}
   \mathbb{I} & 0 \\  \hline
   0 & \begin{smallmatrix} 
    \cos(\theta) & \text{e}^{i\phi}\sin(\theta)^{\phantom{I^2}} \\
    \text{e}^{-i\phi}\sin(\theta) & -\cos(\theta)
  \end{smallmatrix} \\
 \end{array}\right),
\end{align}
where $\mathbb{I}$ is the $6 \times 6$ identity matrix and the superscript c indicates that this is the controlled version of the holonomic gate. We now apply this transformation twice:
\begin{align}
  U^\text{c}(\pi/2,\theta)U^\text{c}(0,0) = \left(\begin{array}{c | c}
   \mathbb{I} & 0 \\ \hline
   0 & \begin{smallmatrix}
    \cos(\theta) & -i\sin(\theta)^{\phantom{I}} \\
    -i\sin(\theta) & \cos(\theta)
  \end{smallmatrix} \\
 \end{array}\right).
\end{align}
This is equal to the famous Deutsch gate except for a factor $i$ on the $2\times2$ rotation matrix. The Deutsch gate is universal for quantum computation \cite{Deutsch73}, and thus our double-controlled holonomic gate is also universal. Implementation of this gate have previously been proposed using Rydberg atoms \cite{PhysRevApplied.9.051001}, albeit using three laser pulses instead of only two as in our case.

In Figure \ref{fig:holonomic_dynamics} we show the evolution of different
initial states under gate operation. Evidently, the qutrit rotation is blocked when the left qubit is in the spin-down state or both qubits are in the spin down state while the qutrit is rotated according to Equation \eqref{eq:holonomic} when both qubits are in the excited state.

\begin{figure}
  \includegraphics[width=0.99\columnwidth]{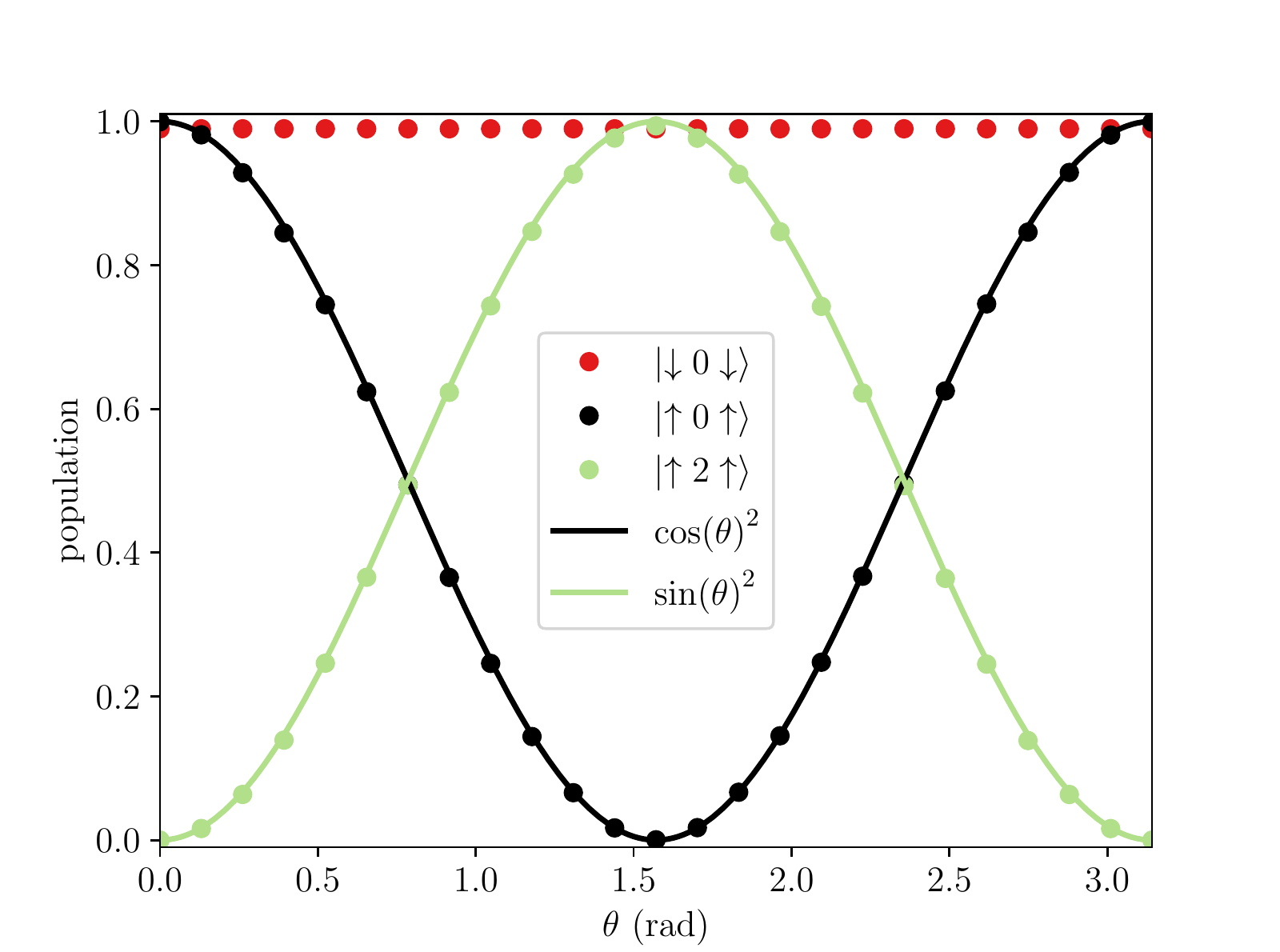}
  \caption{Populations of state versus $\theta$ ($\phi = 0$) after the application of the controlled-controlled holonomic gate to the initial state $\ket{\uparrow 0 \uparrow}$ (black and green points) and $\ket{\downarrow 0 \uparrow}$ (red points). See Supplementary Material for the parameters used in the simulation.}
  \label{fig:holonomic}
\end{figure}

In Figure \ref{fig:holonomic}, we show the populations of the final states
for various values of $\theta$, while $\phi = 0$. The theoretical curves
$\cos^2\!\theta$ and $\sin^2\!\theta $ from \eqref{eq:holonomic} are also shown
and we observe a very good agreement. The final population of the blocked
state $ \ket{\downarrow 0 \uparrow} $ is somewhat lower than expected primarily due to leakage to the other levels via a weak interaction with the external field,
even though it is far from resonance. This leakage is also apparent in
Figure \ref{fig:holonomic_dynamics} and can potentially be reduced by employing pulse shaping techniques \cite{PhysRevLett.103.110501_DRAG}.


\section*{Discussion} \label{sec:discussion}

\noindent To summarize, we have proposed a realistic superconducting circuit, consisting of a qutrit and two qubits, for efficient implementations of multi-qubit quantum gates. By utilizing the second excited state of the qutrit in the middle position, we proposed simple schemes for generating a maximally entangled Bell state of the outer qubits 
and a GHZ state of the qubits and the qutrit. Furthermore, our construction can implement several important quantum gates, such as the \textsc{ccnot} (Toffoli), and \textsc{cswap} (Fredkin) gates. We note that with qubits only, the theoretically most efficient realizations of the Fredkin and Toffoli gates each requires five two-qubit gates \cite{PhysRevA.88.010304_toffoli_minimum}. State of the art implementations of the Toffoli and CCZ gates using superconducting circuits have operation times ranging from $\SI{90}{\nano\second} $ (with poor fidelity) \cite{implementation_toffoli} to about $\SI{260}{\nano\second}$ \cite{PhysRevB.96.024504_toffoli_2017}. As for the Fredkin gate, we are not aware of an implementation with a superconducting circuit, but a hybrid scheme proposal has a gate execution time of $\SI{350}{\nano\second} $ \cite{Liu:18_hybrid_fredkin}. Using the current state of the art superconducting systems requiring $\SI{40}{\nano\second} $ per two-qubit gate \cite{2019arXiv190302492R_fast_two_qubit,PhysRevLett.109.060501_IBM_two_qubit}, the total three-qubit gate time would be at least $ \SI{200}{\nano\second} $. For comparison, our proposed scheme can complete the three-qubit operations in $\SI{100}{\nano\second} $. Thus, our results exemplify the flexibility and usefulness of qutrits for very efficient realizations of three-qubit gates, and demonstrate the potential of our circuit to serve as a basis for more complicated superconducting circuits. 

Our scheme can implement in principle any controlled-controlled unitary operation on the qutrit. As an example, we have considered the double-controlled non-abelian holonomic quantum gate on a single qubit, which can be used to implement the three-qubit Deutsch gate in only two operations. This implementation is more effective than current proposals with Rydberg atoms \cite{PhysRevApplied.9.051001}, while we are not aware of an implementation using superconducting circuits.
We have  implemented the holonomic gate as it is robust to parameter noise \cite{PhysRevA.70.042316} stemming from the geometric nature of this gate. The strategy of using such gates is known as holonomic quantum computation (HQC) \cite{HQC} and the universal non-abelian HQC (NHQC) generalization has since been performed by Sjöqvist et al. \cite{Sjoqvist_NHQC}. Here, three bare energy eigenstates are needed and are conveniently provided by the qutrit. A natural next step is to try to implement the two-qubit non-adiabatic holonomic quantum gate also suggested by Sjöqvist et al., requiring two nearest-neighbor qutrits. Such a gate could be possibly achieved by our circuit upon expanding the basis of one of the outer qubits. Together with the holonomic one-qubit gate, this would realize a universal set of holonomic gates.

Realizing qutrit-qutrit interactions would also open the possibility of implementing higher-order effective spin chains, such as the spin-one Haldane spin model \cite{PhysRevLett.50.1153_Haldane_original,Haldane_spin_gap_review}, especially if the coherence times of higher levels are further prolonged \cite{noise_higher_level}.

Another possible use of qutrits and a circuit similar to the one proposed in this paper is the implementation of autonomous quantum error correction via engineered dissipation. With a relatively small increase in circuit complexity including three energy levels, an impressive increase in transmon coherence time was predicted in Ref. \cite{PhysRevLett.116.150501_Kapit}.

\section*{Methods} \label{subsec:methods}

\noindent Consider the circuit with four connected superconducting islands with lumped element circuit shown in Figure \ref{fig:circuit} (b). After obtaining the Lagrangian of the corresponding effective lumped element model system in the node flux picture, we perform a suitable change of coordinates, primarily mixing the two central flux node coordinates:
$ \psi_1 = \phi_a + \phi_b - 2\phi_c $, $ \psi_2 = \phi_a - \phi_b $ and $ \psi_3 = \phi_a + \phi_b - 2\phi_d $, where the $\phi$s are the flux node variables shown in the circuit (in natural units). They represent the horizontal dipole mode between the left superconducting island and the two middle islands, the vertical dipole mode between the two middle islands, and the horizontal mode between the right island and the two middle islands, respectively.
With this choice of coordinates, we obtain three effective nodes with the relevant degrees of freedom sequentially coupled via non-linear interactions. We truncate the outer nodes to the lowest two states, obtaining qubits, while for the middle node we instead choose to truncate its Hilbert space to the lowest three energy levels, obtaining a qutrit. All three degrees of freedom are in the transmon limit with the kinetic energy terms being much smaller than the potential energy terms. Finally, by transforming to a rotating frame and making a rotating wave approximation to eliminate the fast oscillating terms, we obtain an effective Hamiltonian for the system of two qubits each coupled to the qutrit (see Supplementary Note 1 for the full derivation). This Hamiltonian is given in \eqref{eq:H_full}.

The drive line terms are added to the non-truncated Lagrangian as externally varied flux nodes and a transformation to an appropriate frame rotating with the external field is performed. This transformation mixes the variables and after additional rotating wave approximations, the desired external part of the Hamiltonian is obtained along with modifications on the energy parameters, i.e. the AC-Stark shifts, which can be used for tuning the qubits and qutrit in and out of resonance.

We simulate the dissipative dynamics of the system 
numerically, with the relaxation and decoherence times set to 
$T_1 = \SI{31}{\micro\second}$ and $T_2 = \SI{35}{\micro\second}$ respectively, 
based on recent studies \cite{noise,noise_higher_level,PhysRevB.86.100506} (we use the Python QuTip package \cite{qutip}, and relaxations are implemented by the simple built-in collapse operator functionality). The parameters of the Hamiltonian used in the numerical simulations are all obtained from realistic experimental circuit parameters, as detailed in Supplementary Note 1, and are listed in Supplementary Note 2 for each implementation.

\subsection*{Data availability}
\noindent The data that support the findings of this study are available from N.T.Z. upon request.

\begin{acknowledgments}
We thank W. D. Oliver, S. Gustavsson, and M. Kj{\ae}rgaard from the Engineering Quantum
Systems Group at MIT for their kind hospitality and for
extended discussion on superconducting circuits. This
work was supported by the Carlsberg Foundation and the
Danish Council for Independent Research under the DFF
Sapere Aude program.
\end{acknowledgments}

\subsection*{Author contributions}
The circuit was designed by L.B.K. and N.J.S.L., and analyzed by L.B.K. and T.B. The numerical calculations were performed by T.B. with suggestions from D.P., C.K.A., N.J.S.L. and N.T.Z. The initial draft of the paper was written by T.B., D.P. and N.T.Z. All authors contributed to the revisions that lead to the final version.

\bibliography{qubit-qutrit_manuscript_final}

\end{document}


\title{\Huge Supplementary Information}

\date{}

\maketitle

\section*{Supplementary Note 1: Derivation of the Hamiltonian for the circuit}

\begin{figure*}[htbp]
    \centering
    \includegraphics[width=0.8\textwidth]{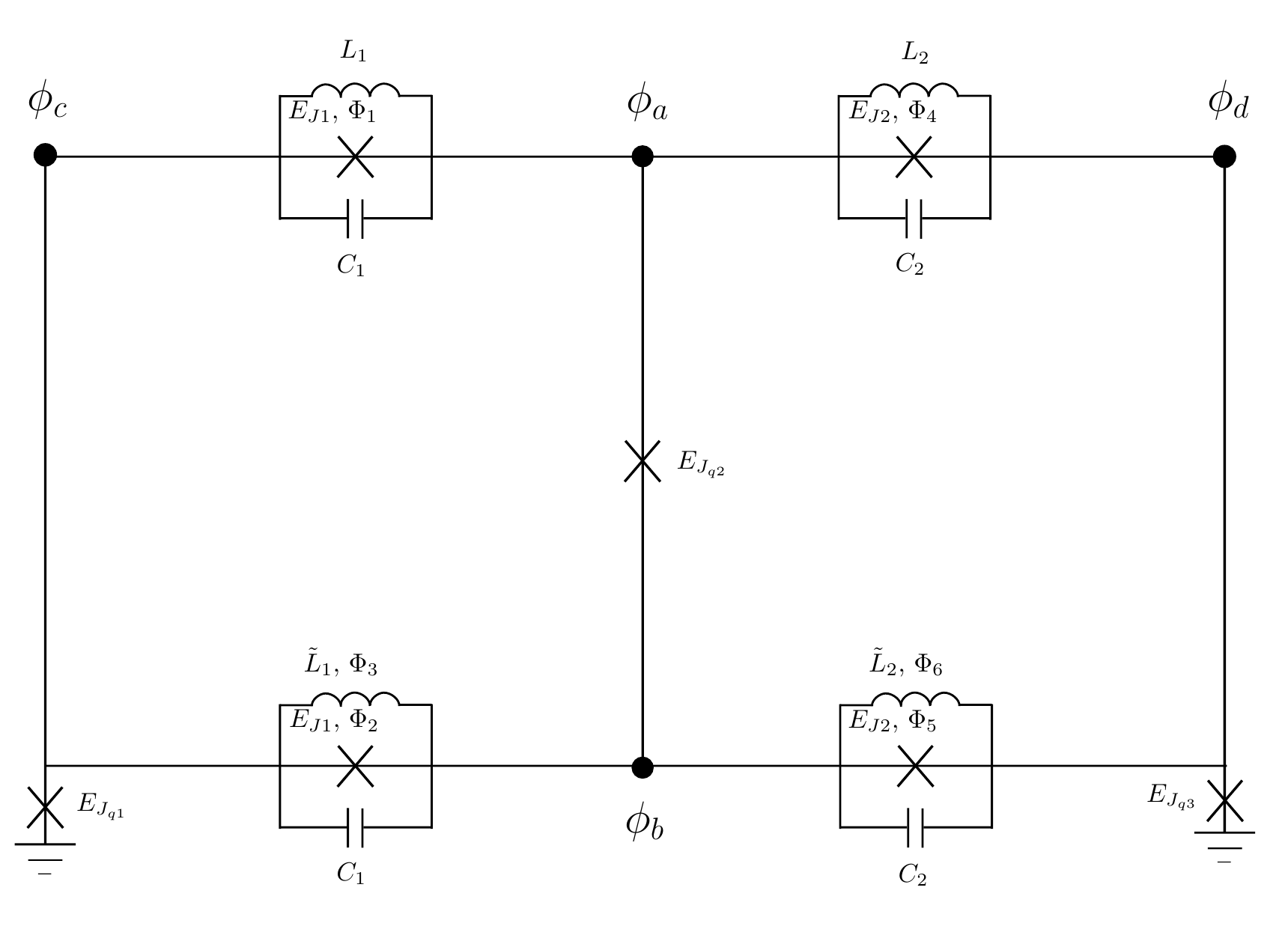}
    \caption{The superconducting circuit and the corresponding parameters describing the properties of the components. Indicated are also the nodes and the corresponding fluxes.}
    \label{fig:big_circuit}
\end{figure*}

We are considering the effective circuit in fig. \ref{fig:big_circuit} and want to show that the low-energy degrees of freedom of this circuit constitutes three qubits(qutrits) with a Heisenberg XXZ-interaction. We start by writing down the Lagrangian of the system in the node flux picture
with the closure branches being the two lower horizontal branches (each splitting into three branches with different circuit elements). The resulting Lagrangian is:
\begin{equation} \label{eq:L_start}
    \begin{split}
        L &= \frac{C_1}{2}\left( \dot{\phi}_a - \dot{\phi}_c \right)^2 + \frac{C_1}{2}\left( \dot{\phi}_b - \dot{\phi}_c \right)^2 - \frac{1}{2L_1}\left( \phi_a - \phi_c \right)^2 - \frac{1}{2\tilde{L}_1}\left( \phi_b - \phi_c + \Phi_3 \right)^2 \\
        &\quad + \frac{C_2}{2}\left( \dot{\phi}_a - \dot{\phi}_d \right)^2 + \frac{C_2}{2}\left( \dot{\phi}_b - \dot{\phi}_d \right)^2 - \frac{1}{2L_2}\left( \phi_a - \phi_d \right)^2 - \frac{1}{2\tilde{L}_2}\left( \phi_b - \phi_d + \Phi_6 \right)^2 \\
        & \quad + E_{J1}\big[ \cos(\phi_a - \phi_c + \Phi_1) + \cos(\phi_b - \phi_c + \Phi_2) \big] \\
        & \quad + E_{J2}\big[ \cos(\phi_a - \phi_d + \Phi_4) + \cos(\phi_b - \phi_d + \Phi_5) \big] \\
        & \quad + E_{J_{q1}}\cos(\phi_c) + E_{J_{q2}}\cos(\phi_a - \phi_b) + E_{J_{q3}}\cos(\phi_d).
    \end{split}
\end{equation}
Defining $\Phi_{\Sigma1} = \Phi_1 + \Phi_2 $ and assuming $ \Phi_1 - \Phi_2 = 0 $, we can rewrite the third line using trigonometric identities:
\begin{equation} \label{eq:using_trig_identities}
    2E_{J1}\cos\left( \frac{\phi_a + \phi_b - 2\phi_c + \Phi_{\Sigma1}}{2} \right) \cos \left( \frac{\phi_a - \phi_b}{2} \right).
\end{equation}
We will stay in the transmon regime 
, where the potential terms are much larger than the kinetic terms. This means that we can assume to be close to the potential minimum, approximated to first order by a harmonic oscillator. Thus, we will later rewrite the Hamiltonian in terms of the bosonic step operators related to the harmonic part of the Hamiltonian. We thereafter employ a rotating wave approximation, removing all terms with odd dependence of the node flux variables since these will be energy non-conserving. Specifically these will, after the truncation to the lowest energy levels, represent spontaneous excitation terms.
From the Taylor expansion of trigonometric functions, we notice that we can further simplify the expression \eqref{eq:using_trig_identities} to the following form using more trigonometric identities:
\begin{equation}
    2E_{J1}\cos\left( \frac{\Phi_{\Sigma1}}{2} \right)\cos\left( \frac{\phi_a + \phi_b - 2\phi_c}{2} \right)\cos\left( \frac{\phi_a - \phi_b}{2} \right).
\end{equation}
Making the same kind of definition for the fourth term, i.e. $ \Phi_{\Sigma2} = \Phi_4 + \Phi_5 $ and assuming $ \Phi_4 - \Phi_5 = 0 $, we can make the same simplification here. We can also ignore the terms dependent on $ \Phi_3 $ and $\Phi_6$ as these will again only be irrelevant offset terms or have an odd dependence on the node fluxes. The Lagrangian becomes
\begin{equation*}
    \begin{split}
        L &= \frac{C_1}{2}\left( \dot{\phi}_a - \dot{\phi}_c \right)^2 + \frac{C_1}{2}\left( \dot{\phi}_b - \dot{\phi}_c \right)^2 - \frac{1}{2L_1}\left( \phi_a - \phi_c \right)^2 - \frac{1}{2\tilde{L}_1}\left( \phi_b - \phi_c \right)^2 \\
        &\quad + \frac{C_2}{2}\left( \dot{\phi}_a - \dot{\phi}_d \right)^2 + \frac{C_2}{2}\left( \dot{\phi}_b - \dot{\phi}_d \right)^2 - \frac{1}{2L_2}\left( \phi_a - \phi_d \right)^2 - \frac{1}{2\tilde{L}_2}\left( \phi_b - \phi_d \right)^2 \\
        & \quad + 2E_{J1}\cos\left( \frac{\Phi_{\Sigma1}}{2} \right)\cos\left( \frac{\phi_a + \phi_b - 2\phi_c}{2} \right)\cos\left( \frac{\phi_a - \phi_b}{2} \right) \\
        & \quad + 2E_{J2}\cos\left( \frac{\Phi_{\Sigma2}}{2} \right)\cos\left( \frac{\phi_a + \phi_b - 2\phi_d}{2} \right)\cos\left( \frac{\phi_a - \phi_b}{2} \right) \\
        & \quad + E_{J_{q1}}\cos(\phi_c) + E_{J_{q2}}\cos(\phi_a - \phi_b) + E_{J_{q3}}\cos(\phi_d).
    \end{split}
\end{equation*}
The next step is defining a suitable set of variables. Inspired by the way the $\phi_i$'s enter the cosines above, we choose the following:
\begin{equation}
    \begin{split}
        \psi_1 &= \phi_a + \phi_b - 2\phi_c \\
        \psi_2 &= \phi_a - \phi_b\\
        \psi_3 &= \phi_a + \phi_b - 2\phi_d \\
        \psi_\text{CM} &= \phi_a + \phi_b\ .
    \end{split}
\end{equation}
In terms of the new variables after expansion of the brackets and collection of terms, the Lagrangian is:
\begin{align*}
    L &= \frac{C_1}{4} {\dot{\psi}_1}^{\ 2} - \left( \frac{1}{8L_1} + \frac{1}{8\tilde{L}_1} \right){\psi_1}^2 + E_{J_{q1}}\cos\left( \frac{\psi_1 - \psi_{CM}}{2} \right) \\
    &\quad + \left( \frac{C_1}{4} + \frac{C_2}{4} \right)\dot{\psi}_2^{\ 2} - \left( \frac{1}{8L_1} + \frac{1}{8\tilde{L}_1} + \frac{1}{8L_2} + \frac{1}{8\tilde{L}_2} \right){\psi_2}^2 + E_{J_{q2}} \cos\left( \psi_2 \right)\\
    &\quad + \frac{C_2}{4}{\dot{\psi}_3}^{\ 2} - \left( \frac{1}{8L_2} + \frac{1}{8\tilde{L}_2} \right){\psi_3}^2 + E_{J_{q3}}\cos\left( \frac{\psi_3 - \psi_{CM}}{2} \right) \\
    &\quad - \left( \frac{1}{8L_1} - \frac{1}{8\tilde{L}_1} \right)\psi_1\psi_2 + 2E_{J1}\cos\left( \frac{\Phi_{\Sigma1}}{2} \right)\cos\left( \frac{\psi_1}{2} \right)\cos\left( \frac{\psi_2}{2} \right)\\
    &\quad - \left( \frac{1}{8L_2} - \frac{1}{8\tilde{L}_2} \right)\psi_2\psi_3 + 2E_{J2}\cos\left( \frac{\Phi_{\Sigma2}}{2} \right)\cos\left( \frac{\psi_2}{2} \right)\cos\left( \frac{\psi_3}{2} \right).
\end{align*}
The conjugate momenta can be found from the usual definition $p_i = \frac{\partial L}{\partial \dot{\psi}_i} $:
\begin{equation}
    \begin{split}
        p_1 &= \frac{C_1}{2}\dot{\psi}_1 \\
        p_2 &= \left( \frac{C_1}{2} + \frac{C_2}{2} \right)\dot{\psi}_2 \\
        p_3 &= \frac{C_2}{2}\dot{\psi}_3 \\
        p_{_{CM}} &= 0 \ .
    \end{split}
\end{equation}
Notice that the conjugate momentum of $\psi_{CM}$ is zero, and thus the variable can be seen as purely a constraint variable without a kinetic term. We will thus ignore $\psi_{CM}$ from now on.

Now performing a Legendre transforming to the Hamiltonian via the relation $ H = \sum_i \dot{\psi}_i p_i - L $, we obtain:
\begin{align*}
    H &= \frac{1}{C_1}{p_1}^2 + \left( \frac{1}{8L_1} + \frac{1}{8\tilde{L}_1} \right){\psi_1}^2 - E_{J_{q1}}\cos\left( \frac{\psi_1}{2} \right) \\
    &\quad + \frac{1}{C_1+C_2}{p_2}^2 + \left( \frac{1}{8L_1} + \frac{1}{8\tilde{L}_1} + \frac{1}{8L_2} + \frac{1}{8\tilde{L}_2} \right){\psi_2}^2 - E_{J_{q2}} \cos\left( \psi_2 \right)\\
    &\quad + \frac{1}{C_2}{p_3}^2 + \left( \frac{1}{8L_2} + \frac{1}{8\tilde{L}_2} \right){\psi_3}^2 - E_{J_{q3}}\cos\left( \frac{\psi_3}{2} \right) \\
    &\quad + \left( \frac{1}{8L_1} - \frac{1}{8\tilde{L}_1} \right)\psi_1\psi_2 - 2E_{J1}\cos\left( \frac{\Phi_{\Sigma1}}{2} \right)\cos\left( \frac{\psi_1}{2} \right)\cos\left( \frac{\psi_2}{2} \right)\\
    &\quad + \left( \frac{1}{8L_2} - \frac{1}{8\tilde{L}_2} \right)\psi_2\psi_3 - 2E_{J2}\cos\left( \frac{\Phi_{\Sigma2}}{2} \right)\cos\left( \frac{\psi_2}{2} \right)\cos\left( \frac{\psi_3}{2} \right).
\end{align*}
The cosines are now expanded to fourth order in the flux variables:
\begin{equation}
    \begin{split}
        \cos(x) &\simeq 1 - \frac{1}{2}x^2 + \frac{1}{24}x^4 \\
        \cos\left( \frac{x}{2} \right) &\simeq 1 - \frac{1}{8}x^2 + \frac{1}{384}x^4 \\
        \cos\left( \frac{x_1}{2} \right)\cos\left( \frac{x_2}{2} \right) &\simeq 1 - \frac{1}{8}{x_1}^2 + \frac{1}{384}{x_1}^4 - \frac{1}{8}{x_2}^2 + \frac{1}{384}{x_2}^4 + \frac{1}{64}{x_1}^2{x_2}^2\ .
    \end{split}
\end{equation}
This approximation neglects six-order terms and higher. Using first-order perturbation theory on the sixth-order term from $\cos\left( \psi_2 \right)$ above, we get a correction to the final ground state energy of the corresponding qubit of $ E^{(1)}/(2\pi) \simeq \SI{31}{\kilo\hertz} $. This can safely be ignored compared to the usual transmon energies in the order of $\SI{10}{\giga\hertz} $. Also, since we make sure to stay in the transmon limit with the potential terms much higher than the kinetic terms, we will always stay near very small values of the transmon phase drops, which is the physical interpretation of the flux variables, $\psi_i$. This makes, up to irrelevant constant terms,

\begin{align*}
    H &= \frac{1}{C_1}{p_1}^2 + \alpha_1 {\psi_1}^2 - \beta_1 {\psi_1}^4 \\
    &\quad +\frac{1}{C_1+C_2}{p_2}^2 + \alpha_2{\psi_2}^2 - \beta_2 {\psi_2}^4 \\
    &\quad +\frac{1}{C_2}{p_3}^2 + \alpha_3{\psi_3}^2 - \beta_3{\psi_3}^4 \\
    &\quad +\left( \frac{1}{8L_1} - \frac{1}{8\tilde{L}_1} \right)\psi_1\psi_2 - \frac{E_{J1}}{32}\cos\left( \frac{\Phi_{\Sigma1}}{2} \right){\psi_1}^2 {\psi_2}^2 \\
    &\quad +\left( \frac{1}{8L_2} - \frac{1}{8\tilde{L}_2} \right)\psi_2\psi_3 - \frac{E_{J2}}{32}\cos\left( \frac{\Phi_{\Sigma2}}{2} \right){\psi_2}^2 {\psi_3}^2,
\end{align*}
where
\begin{equation} \label{eq:alpha_beta}
    \begin{split}
        \alpha_1 &= \frac{1}{8L_1} + \frac{1}{8\tilde{L}_1} + \frac{E_{J_{q1}}}{8} + \frac{E_{J1}}{4}\cos\left( \frac{\Phi_{\Sigma1}}{2} \right) \\
        \beta_1 &= \frac{E_{J_{q1}}}{384} + \frac{E_{J1}}{192}\cos\left( \frac{\Phi_{\Sigma1}}{2} \right)\\
        \alpha_2 &= \frac{1}{8L_1} + \frac{1}{8\tilde{L}_1} + \frac{1}{8L_2} + \frac{1}{8\tilde{L}_2} + \frac{E_{J_{q2}}}{2} + \frac{E_{J1}}{4}\cos\left( \frac{\Phi_{\Sigma1}}{2} \right) + \frac{E_{J2}}{4}\cos\left( \frac{\Phi_{\Sigma2}}{2} \right) \\
        \beta_2 &= \frac{E_{J_{q2}}}{24} + \frac{E_{J1}}{192}\cos\left( \frac{\Phi_{\Sigma1}}{2} \right) + \frac{E_{J2}}{192}\cos\left( \frac{\Phi_{\Sigma2}}{2} \right) \\
        \alpha_3 &= \frac{1}{8L_2} + \frac{1}{8\tilde{L}_2} + \frac{E_{J_{q3}}}{8} + \frac{E_{J2}}{4}\cos\left( \frac{\Phi_{\Sigma2}}{2} \right) \\
        \beta_3 &= \frac{E_{J_{q3}}}{384} + \frac{E_{J2}}{192}\cos\left( \frac{\Phi_{\Sigma2}}{2} \right).
    \end{split}
\end{equation}
Taking the anharmonicity (fourth-order terms) as a perturbation, we now choose the bosonic raising and lowering operators related to the harmonic parts of the Hamiltonian (the $p^2$ and $\psi^2$ -terms) above as usual (This perturbation is essential for the later truncation, since the resulting anharmonicity of the energy spacing between the levels allows us to only consider the lowest states):
\begin{equation}
    \begin{split}
        \psi_1 &= \frac{1}{(4\alpha_1C_1)^\frac{1}{4}}\left( b_1^\dagger + b_1 \right) \\
        \psi_2 &= \frac{1}{(4\alpha_2(C_1+C_2))^\frac{1}{4}}\left( b_2^\dagger + b_2 \right) \\
        \psi_3 &= \frac{1}{(4\alpha_3C_2)^\frac{1}{4}}\left( b_3^\dagger + b_3 \right).
    \end{split}
\end{equation}
We end up with the rewritten Hamiltonian
\begin{equation}
	\begin{aligned} \label{eq:H_b}
    H &= \sqrt{\frac{4\alpha_1}{C_1}}\left( b_1^\dagger b_1 + \frac{1}{2} \right) - \frac{\beta_1}{4\alpha_1C_1}\left( b_1^\dagger + b_1 \right)^4 \\
    &\quad + \sqrt{\frac{4\alpha_2}{C_1+C_2}}\left( b_2^\dagger b_2 + \frac{1}{2} \right) - \frac{\beta_2}{4\alpha_2(C_1+C_2)}\left( b_2^\dagger + b_2 \right)^4 \\
    &\quad +\sqrt{\frac{4\alpha_3}{C_2}}\left( b_3^\dagger b_3 + \frac{1}{2} \right) - \frac{\beta_3}{4\alpha_3C_2}\left( b_3^\dagger + b_3 \right)^4 \\
    &\quad + \left( \frac{1}{8L_1} - \frac{1}{8\tilde{L}_1} \right)\frac{1}{2(\alpha_1\alpha_2C_1(C_1+C_2))^\frac{1}{4}}\left( b_1^\dagger + b_1 \right)\left( b_2^\dagger + b_2 \right) \\
    &\quad - \frac{E_{J1}}{128}\cos\left( \frac{\Phi_{\Sigma1}}{2} \right)\frac{1}{\sqrt{\alpha_1\alpha_2C_1(C_1+C_2)}}\left( b_1^\dagger + b_1 \right)^2 \left( b_2^\dagger + b_2 \right)^2\\
    &\quad + \left( \frac{1}{8L_2} - \frac{1}{8\tilde{L}_2} \right)\frac{1}{2(\alpha_2\alpha_3C_2(C_1+C_2))^\frac{1}{4}}\left( b_2^\dagger + b_2 \right)\left( b_3^\dagger + b_3 \right) \\
    &\quad - \frac{E_{J2}}{128}\cos\left( \frac{\Phi_{\Sigma2}}{2} \right)\frac{1}{\sqrt{\alpha_2\alpha_3C_2(C_1+C_2)}}\left( b_2^\dagger + b_2 \right)^2 \left( b_3^\dagger + b_3 \right)^2.
\end{aligned}
\end{equation}
We could now in principle map this to a spin model by using the anharmonicity to truncate to the lowest two/three energy eigenstates (e.g. $ b_i^\dagger + b_i \mapsto \sigma_i^x$ for a qubit) and after using the rotating wave approximation end up with a Hamiltonian with the desired form. This would have the drawback of being static, i.e. once the circuit is built to certain specifications the experimental parameters are fixed and the energy levels and coupling strengths can not be adjusted significantly afterwards. We will introduce a dynamical tuning by adding an external driving field, effectively mixing the first and second excited state of the middle ($\psi_2$) degree of freedom and thereby changing the energy levels and coupling strengths.

\subsection{Adding an effective external energy level tuning}

We now imagine control lines connecting to the nodes $a$ and $b$ as in the main text. These can then be used to drive the middle degree of freedom $\psi_2$ using external fields. Specifically, we let the control line 5 connect to the node with flux $\phi_a$ in Supplemetary Fig. \ref{fig:big_circuit} through the capacitance $C_\text{ext}$ and drive an external field $\phi_\text{ext}$. Similarly, we apply an external field $\phi_{-\text{ext}} $ through control line 6 connected to the node flux $\phi_b $ through the same capacitance $C_\text{ext} $.
The following extra term will appear in the Lagrangian:
\begin{equation}
    L_\text{ext} = \frac{C_\text{ext}}{2}\left( \dot{\phi}_a - \dot{\phi}_\text{ext} \right)^2 + \frac{C_\text{ext}}{2}\left( \dot{\phi}_b - \dot{\phi}_{-\text{ext}} \right)^2 .
\end{equation}
This can be rewritten to
\begin{equation*}
    L_\text{ext} = \frac{C_\text{ext}}{4}\left[ \left( \dot{\phi}_a + \dot{\phi}_b - \dot{\phi}_{-\text{ext}} - \dot{\phi}_\text{ext} \right)^2 + \left( \dot{\phi}_a - \dot{\phi}_b + \dot{\phi}_{-\text{ext}} - \dot{\phi}_\text{ext} \right)^2 \right].
\end{equation*}
Assuming $\phi_\text{ext} = - \phi_{-\text{ext}} = A_\text{ext} \sin(\omega_\text{ext} t)$ and transforming to the $\psi$-coordinates, this reduces to
\begin{align}
    L_\text{ext} &= \frac{C_\text{ext}}{4}\left[\dot{\psi}_{CM}^{\quad 2} + \left( \dot{\psi}_2 - 2A_\text{ext}\omega_\text{ext}\cos(\omega_\text{ext} t) \right)^2\right] \nonumber\\
    &= \frac{C_\text{ext}}{4}\dot{\psi}_{CM}^{\quad 2} + \frac{C_\text{ext}}{4}{\dot{\psi}_2}^2 + C_\text{ext} A_\text{ext}^2 \omega_\text{ext}^2\cos(\omega_\text{ext} t)^2 - C_\text{ext} \dot{\psi}_2 A_\text{ext} \omega_\text{ext} \cos(\omega_\text{ext} t).
\end{align}
The first term here is an apparently problematic kinetic term for what has so far been a constraint variable. This can, though, be constructed to have a spacing between the energy levels that is very far from the spacing of the other degrees of freedom and $ \psi_{CM} $ can thus be ignored in spite of this term.
The second term is another kinetic term for the $\psi_2$-variable, the third term is an offset term, and the fourth and last term is an interaction term between $\psi_2$ and the external field.
It is this last term that will be useful for driving transitions between, and hence coupling of, the first and second excited levels of $\psi_2$. This will allow us to tune the position of the two levels through non-crossing depending on how much we mix the states.

Including the above addition, the Lagrangian related purely to the $\psi_2$-degree of freedom is the following:
\begin{equation*}
    L_2 = \frac{C_1 + C_2 + C_\text{ext}}{4}\dot{\psi}_2^{\ 2} - \left( \frac{1}{8L_1} + \frac{1}{8\tilde{L}_1} + \frac{1}{8L_2} + \frac{1}{8\tilde{L}_2} \right){\psi_2}^2 + E_{J_{q2}}\cos(\psi_2) - C_\text{ext} A_\text{ext} \omega_\text{ext} \cos(\omega_\text{ext} t)\dot{\psi}_2,
\end{equation*}
with all other terms being either purely related to $\psi_1$ or $\psi_3$ or interaction terms.
The corresponding conjugated momentum is also altered slightly:
\begin{equation}
    p_2 = \frac{C_1+C_2+C_\text{ext}}{2}\dot{\psi}_2 - C_\text{ext} A_\text{ext} \omega_\text{ext} \cos(\omega_\text{ext} t),
\end{equation}
thus
\begin{equation}
    \dot{\psi}_2 = \frac{2}{C_1 + C_2+C_\text{ext}}\big( p_2 + C_\text{ext} A_\text{ext} \omega_\text{ext} \cos(\omega_\text{ext} t) \big).
\end{equation}
The related Hamiltonian is therefore, ignoring any offset terms:
\begin{align*}
    H_2 &= \frac{{p_2}^2}{C_1+C_2+C_\text{ext}} + \frac{2C_\text{ext} A_\text{ext} \omega_\text{ext} \cos(\omega_\text{ext} t)}{C_1+C_2+C_\text{ext}}p_2 \\
    &\quad + \left( \frac{1}{8L_1} + \frac{1}{8\tilde{L}_1} + \frac{1}{8L_2} + \frac{1}{8\tilde{L}_2} \right){\psi_2}^2 - E_{J_{q2}}\cos(\psi_2).
\end{align*}
Performing the same expansions as before addition of the extra capacitances and external fields gives a few extra $\psi_2$-terms from the interaction terms. Focusing on only the resulting terms containing only $\psi_2$ or $p_2$, we get:
\begin{equation}
    H_2 = \frac{1}{C_1 + C_2 + C_\text{ext}}{p_2}^2 + \alpha_2 {\psi_2}^2 - \beta_2 {\psi_2}^4 + \frac{2C_\text{ext} A_\text{ext} \omega_\text{ext} \cos(\omega_\text{ext} t)}{C_1 + C_2 + C_\text{ext}}p_2,
\end{equation}
where $\alpha_2$ and $\beta_2$ are defined in \eqref{eq:alpha_beta}. We once again introduce the bosonic step-operators related to the harmonic part of $H_2$:
\begin{equation}
    \begin{split}
        \psi_2 &= \frac{1}{(4\alpha_2(C_1+C_2+C_\text{ext}))^\frac{1}{4}}\left( b_2^\dagger + b_2 \right) \\
        p_2 &= i\frac{(4\alpha_2(C_1+C_2+C_\text{ext}))^\frac{1}{4}}{2} \left( b_2^\dagger - b_2 \right),
    \end{split}
\end{equation}
which makes
\begin{equation} \label{eq:H_2_plus_external}
    H_2 = T_1 b_2^\dagger b_2 - T_2\left( b_2^\dagger + b_2 \right)^4 + iT_\text{ext}\cos \left( \omega_\text{ext} t \right)\left( b_2^\dagger - b_2 \right),
\end{equation}
where
\begin{equation}
    \begin{split}
        T_1 &= \sqrt{\frac{4\alpha_2}{C_1+C_2+C_\text{ext}}} \\
        T_2 &= \frac{\beta_2}{4\alpha_2(C_1+C_2+C_\text{ext})} \\
        T_\text{ext} &= C_\text{ext} A_\text{ext} \omega_\text{ext} \frac{(4\alpha_2(C_1 + C_2 + C_\text{ext}))^{\frac{1}{4}}}{C_1 + C_2 + C_\text{ext}}.
    \end{split}
\end{equation}
We wish to diagonalize this Hamiltonian to investigate the effect of $T_\text{ext}$ on the spectrum.

\subsection{Truncating the middle degree of freedom}

We will now investigate the dynamical tuning of the spectrum by first truncating the ``internal'' Hamiltonian for the $\psi_2$ degree of freedom, corresponding to setting $A_\text{ext} = 0$, to the three lowest degrees of freedom, i.e. find the qutrit eigenstates. We will then add the external field and transform to a frame rotating with the external field, wherein we can see the effective mixing of the qutrit eigenstates. Lastly, we shift basis and use these driven states in the rotating frame as our qutrit eigenstates.

We start by diagonalizing the internal Hamiltonian of the middle degree of freedom ``$H_{2,0}$'', i.e. the first two terms in \eqref{eq:H_2_plus_external}:
In the basis of the three lowest simple harmonic oscillator states, which is chosen since we wish to end up with a qutrit in the end, we represent (up to irrelevant offset terms proportional to the identity)
\begin{align}\label{eq:b_bd}
     b_2^\dagger \sim \begin{pmatrix}
     	0 & 0 & 0 \\ 1 & 0 & 0 \\ 0 & \sqrt{2} & 0
     \end{pmatrix}, \qquad b_2 \sim \begin{pmatrix}
     	0 & 1 & 0 \\ 0 & 0 & \sqrt{2} \\ 0 & 0 & 0
     \end{pmatrix}.
\end{align}
This truncation is also done for all terms in \eqref{eq:H_2_plus_external}, after using the canonical commutation relation $[b,b^\dagger]=1$ to transform to normal ordering form:
\begin{align*}
    b_2^\dagger b_2 \sim \begin{pmatrix}
    	0 & 0 & 0 \\ 0 & 1 & 0 \\ 0 & 0 & 2
    \end{pmatrix} \quad, \qquad \left( b_2^\dagger + b_2 \right)^4 \sim \begin{pmatrix}
    	0 & 0 & 6\sqrt{2} \\ 0 & 12 & 0 \\ 6\sqrt{2} & 0 & 36
    \end{pmatrix},
\end{align*}
and
\begin{align*}
    b_2^\dagger - b_2 \sim \begin{pmatrix}
        0 & -1 & 0 \\
        1 & 0 & -\sqrt{2} \\
        0 & \sqrt{2} & 0
    \end{pmatrix}
\end{align*}
Inserting this in the first two terms in equation \eqref{eq:H_2_plus_external}, we get the Hamiltonian
\begin{align*}
    H_{2,0} \sim \begin{pmatrix}
    	0 & 0 & -6\sqrt{2}T_2 \\ 0 & T_1 -12T_2 & 0 \\ -6\sqrt{2}T_2 & 0 & 2T_1 - 36T_2
    \end{pmatrix},
\end{align*}
which has the eigenenergies:
\begin{equation}
\begin{split}
    E_0 &= T_1 - 18T_2 - \sqrt{(T_1 - 18T_2)^2 + 72 {T_2}^2} \\
    E_1 &= T_1 - 12T_2 \\
    E_2 &= T_1 - 18T_2 + \sqrt{(T_1 - 18T_2)^2 + 72 {T_2}^2}
\end{split}
\end{equation}
with the corresponding eigenstates:
\begin{equation}
\begin{split}
    \ket{\tilde{0}} &= \frac{1}{\sqrt{72{T_2}^2 + {E_0}^2}} \begin{pmatrix} 6\sqrt{2}T_2 \\ 0 \\ -E_0 \end{pmatrix} \\
    \ket{\tilde{1}} &= \begin{pmatrix} 0 \\ 1 \\ 0 \end{pmatrix} \\
    \ket{\tilde{2}} &= \frac{1}{\sqrt{72{T_2}^2 + {E_2}^2}} \begin{pmatrix} -6\sqrt{2}T_2 \\ 0 \\ E_2 \end{pmatrix}.
\end{split}
\end{equation}
Now turning on the full external field, the full Hamiltonian $H_2$ can be expressed as
\begin{align}\label{eq:H_2_tilde_nonrotating}
    H_2 = E_0 \ketbra{\tilde{0}} + E_1 \ketbra{\tilde{1}} + E_2 \ketbra{\tilde{2}} + \frac{T_\text{ext}}{2}\left( \e{i\omega_\text{ext} t} + \e{-i\omega_\text{ext} t} \right) \begin{pmatrix}
        0 & -i & 0 \\
        i & 0 & -\sqrt{2}i \\
        0 & \sqrt{2}i & 0
    \end{pmatrix},
\end{align}
where the first three terms are the bare qutrit eigenstates and the last matrix is written in the old basis $ (\ket{0},\ket{1},\ket{2}) $.
The outer degrees of freedom ($1,3$) is truncated to the lowest two degrees levels (i.e. an effective qubit) as is standard, where $ \ket{\uparrow} $ and $ \ket{\downarrow} $ will denote the excited and ground state, respectively.

We now switch to a rotating frame corresponding to the external field, which means performing a unitary transformation with the operator:
\begin{equation}
    U_{\text{ext}} = U_{\text{ext},1}U_{\text{ext},2}U_{\text{ext},3},
\end{equation}
where
\begin{equation}
    U_{\text{ext},\alpha} =  \text{e}^{i\omega_\text{ext}t/2} \ketbra{\downarrow} + \text{e}^{-i\omega_\text{ext}t/2} \ketbra{\uparrow},
\end{equation}
for $ \alpha = 1,3 $, and
\begin{equation}
    U_{\text{ext},2}(t) = \text{e}^{i3\omega_\text{ext}t/2} \ketbra{\tilde{0}} + \text{e}^{i\omega_\text{ext}t/2} \ketbra{\tilde{1}} + \text{e}^{-i\omega_\text{ext}t/2} \ketbra{\tilde{2}}.
\end{equation}
We are trying to obtain a mixing of the first and second energy levels and therefore assume that we can tune $\omega_\text{ext}$ so that it is close to $E_2 - E_1$ and far from $E_1-E_0$ and $E_2-E_0$, effectively enabling us to perform the two-level approximation.
A transformation of the Hamiltonian can now be performed according to the standard transformation rule
\begin{equation} \label{eq:H_transformation_rotating}
    H \rightarrow H^R = U_{\text{ext}}^\dagger H\, U_{\text{ext}} + i \dv{U_{\text{ext}}^\dagger}{t} U_{\text{ext}}.
\end{equation}
Since our Hamiltonian is quite big, we take this one part at a time. Starting with the terms purely related to the $\psi_2$ degree of freedom, it is a good idea to look at the matrix elements from the last factor in \eqref{eq:H_2_tilde_nonrotating} when performing this transformation:
\begin{align}
    \bra{\tilde{1}}\left( \ketbra{1}{0} -\sqrt{2} \ketbra{1}{2} \right)\ket{\tilde{2}}  &= -\frac{\sqrt{2}(6T_2 + E_2)}{\sqrt{(6\sqrt{2}T_2)^2 + {E_2}^2}} \\
    \bra{\tilde{2}}\left( -\ketbra{0}{1} +\sqrt{2} \ketbra{2}{1} \right)\ket{\tilde{1}}  &= \frac{\sqrt{2}(6T_2 + E_2)}{\sqrt{(6\sqrt{2}T_2)^2 + {E_2}^2}},
\end{align}
where the rest are either irrelevant or zero.
So, we get in the rotating frame
\begin{align}
    H_2 &= E_0 \ketbra{ \tilde{0} } + E_1 \ketbra{ \tilde{1} } + E_2 \ketbra{ \tilde{2} } \nonumber \\
    &\quad+ i\frac{T_\text{ext}}{2} \frac{\sqrt{2}(6T_2 + E_2)}{\sqrt{(6\sqrt{2}T_2)^2 + {E_2}^2}} \left(\e{i\omega_\text{ext} t} + \e{-i\omega_\text{ext} t}\right)\left(-\ketbra{\tilde{1}}{\tilde{2}}\e{-i\omega_\text{ext} t} + \ketbra{\tilde{2}}{\tilde{1}}\e{i\omega_\text{ext} t} \right) \nonumber \\
    &\quad + \frac{3\omega_\text{ext}}{2} \ketbra{\tilde{0}} + \frac{\omega_\text{ext}}{2} \ketbra{\tilde{1}} -\frac{\omega_\text{ext}}{2}\ketbra{\tilde{2}} \nonumber \\
    &= \left(E_0+\frac{3\omega_\text{ext}}{2} \right) \ketbra{\tilde{0}} + \left(E_1+\frac{\omega_\text{ext}}{2} \right)\ketbra{ \tilde{1} } + \left(E_2 - \frac{\omega_\text{ext}}{2}\right) \ketbra{ \tilde{2} } \nonumber\\
    & \quad- i\Delta \Big[ \ketbra{\tilde{1}}{\tilde{2}}\left(1+\e{-2i\omega_\text{ext} t}\right) - \ketbra{\tilde{2}}{\tilde{1}}\left(1+\e{2i\omega_\text{ext} t}\right) \Big] \nonumber \\
    &\simeq \left(E_0+\frac{3\omega_\text{ext}}{2} \right) \ketbra{\tilde{0}} + \left(E_1+\frac{\omega_\text{ext}}{2} \right)\ketbra{ \tilde{1} } + \left(E_2 - \frac{\omega_\text{ext}}{2}\right) \ketbra{ \tilde{2} }+ i\Delta \Big( \ketbra{\tilde{2}}{\tilde{1}} - \ketbra{\tilde{1}}{\tilde{2}} \Big),
\end{align}
where we in the last equation has used the rotating wave approximation to remove the fast oscillating terms $\exp(\pm 2i\omega_\text{ext} t)$, and we have defined
\begin{equation}
    \Delta = \frac{T_\text{ext}}{2} \frac{\sqrt{2}(E_2 + 6T_2)}{\sqrt{(6\sqrt{2}T_2)^2 + {E_2}^2}}.
\end{equation}
Writing this in terms of the detuning from resonance
\begin{equation}
    \delta = E_2 - E_1 - \omega_\text{ext},
\end{equation}
we can rewrite the expression above as
\begin{equation}
    H_2^R =  \left(\frac{E_1+E_2}{2} + \xi - \frac{3\delta}{2} \right) \ketbra{\tilde{0}} + \left(\frac{E_1+E_2}{2} - \frac{\delta}{2} \right)\ketbra{ \tilde{1} } + \left(\frac{E_1+E_2}{2} + \frac{\delta}{2} \right) \ketbra{ \tilde{2} }+ i\Delta \Big( \ketbra{\tilde{2}}{\tilde{1}} - \ketbra{\tilde{1}}{\tilde{2}} \Big)
\end{equation}
or, representing this in the matrix representation in the tilde states as basis states:
\begin{align*}
    H_2^R \sim \begin{pmatrix}
    	  \xi - \frac{3\delta}{2} & 0 & 0 \\
        0 & - \frac{\delta}{2} & -i\Delta \\
        0 & i\Delta & \frac{\delta}{2}
    \end{pmatrix} + \frac{E_1+E_2}{2} \mathbb{I}_3,
\end{align*}
where $\mathbb{I}_3 $ is the $3\times3$ identity matrix of the qutrit, which can be safely ignored, and $ \xi = E_2 - E_1 - (E_1 - E_0)  < 0 $ is the absolute anharmonicity between the first and second level in the qutrit.
This can be diagonalized to find the energy spectrum
\begin{equation}
	\begin{aligned}
		E_0' &=  \xi - \frac{3\delta}{2} \\
		E_1' &= - \gamma \\
		E_2' &= + \gamma ,
	\end{aligned}
\end{equation}
where
\begin{align}
    \gamma = \frac{1}{2}\sqrt{\delta^2 + 4\Delta^2}.
\end{align}
The (normalized) eigenstates of this Hamiltonian, expressed in the basis $\{ \ket{\tilde{0}},\ket{\tilde{1}},\ket{\tilde{2}} \}$, are
\begin{equation}
	\begin{aligned}
		\ket{0'} &\sim \begin{pmatrix} 1 \\ 0 \\ 0 \end{pmatrix} \\
		\ket{1'} &\sim \frac{1}{\sqrt{\Delta^2 + \left( \frac{\delta}{2} - \gamma \right)^2 }}\begin{pmatrix} 0 \\ i\Delta \\ -\frac{\delta}{2} + \gamma\end{pmatrix} \\
		\ket{2'} &\sim \frac{1}{\sqrt{\Delta^2 + \left( \frac{\delta}{2} + \gamma \right)^2 }}\begin{pmatrix} 0 \\ -i\Delta \\ \frac{\delta}{2} + \gamma \end{pmatrix}.
	\end{aligned}
\end{equation}
In the limit $ \Delta \rightarrow 0 $, these reduce to the bare energy states $ \ket{i'} \rightarrow \ket{\tilde{i}} $ for $i=0,1,2$ when $\delta > 0$ meaning the driving frequency is below the undriven energy difference between the upper qutrit states.
The Hamiltonian for the $\psi_2$ degree of freedom reduces to
\begin{align}
    H_2^R = E_0' \ketbra{0'} + E_1' \ketbra{1'} + E_2'\ketbra{2'} = E_0'\mathbb{I}_2 + (E_1' - E_0') \ketbra{1'} + \left( E_2' - E_0' \right)\ketbra{2'},
\end{align}
where the diagonal term will again be throw away from this point onwards.
In conclusion, we see that by tuning $A_\text{ext}$ and/or $\omega_\text{ext}$, we can change $E_1'$ and $E_2'$, i.e. the contribution of the part of the Hamiltonian purely related to $\psi_2$ to the energy of the two highest qutrit states.\\

Next, we perform the transformation to the rotating picture to the parts of the Hamiltonian purely related to the outer fluxes $j = 1,3$. We choose the qubit spin-up state as the excited state, e.g. we associate $ b_j^\dagger b_j \mapsto \tfrac{1}{2} + \tfrac{1}{2}\sigma_j^z $. Thus
\begin{align}
    H_j \rightarrow H_j^R &= U_{\text{ext},j}^\dagger \left[ \sqrt{\frac{4\alpha_j}{C_j}}\left( b_j^\dagger b_j + \frac{1}{2} \right) - \frac{\beta_j}{4\alpha_jC_j}\left( b_j^\dagger + b_j \right)^4 \right]  U_{\text{ext},\alpha} \\
    &= \frac{1}{2}\left( \sqrt{\frac{4\alpha_j}{C_j}} - \frac{3\beta_j}{\alpha_jC_j} \right)\sigma_j^z + \frac{\omega_\text{ext}}{2} \ketbra{\downarrow} - \frac{\omega_\text{ext}}{2} \ketbra{\uparrow} \\
    &= \frac{1}{2}\left( \sqrt{\frac{4\alpha_j}{C_j}} - \frac{3\beta_j}{\alpha_jC_j} - \omega_\text{ext} \right) \sigma_j^z,
\end{align}
where we have made the usual truncation to a qubit.
\\

We have yet to do the transformation to the rotating picture for the interaction terms. The factors involved are $( b_j^\dagger + b_j )$, which normally for a qubit maps to $\sigma^x_j$ before moving to the interaction picture, and $( b_j^\dagger + b_j )^2$, which maps to $2+\sigma_j^z$ for a qubit.
We start by looking at the factor (for $j=1,3$):
\begin{align*}
    (b_2^\dagger + b_2)(b_j^\dagger + b_j) &\rightarrow U_{\text{ext},2}^\dagger \left( b_2^\dagger + b_2\right)U_{\text{ext},2} \ U_{\text{ext},j}^\dagger\left(b_j^\dagger + b_j\right)U_{\text{ext},j}.
\end{align*}
The first factor is:
\begin{align}
    U_{\text{ext},2}^\dagger \left( b_2^\dagger + b_2\right)U_{\text{ext},2} = k_0 \left( \e{-i\omega_\text{ext}t}\ketbra{\tilde{0}}{\tilde{1}} + \e{i\omega_\text{ext}t}\ketbra{\tilde{1}}{\tilde{0}} \right) + k_2 \left( \e{-i\omega_\text{ext}t}\ketbra{\tilde{1}}{\tilde{2}} + \e{i\omega_\text{ext}t}\ketbra{\tilde{2}}{\tilde{1}} \right),
\end{align}
where
\begin{align}
    k_i = (-1)^{\frac{i}{2}}\frac{\sqrt{2}(6T_2 - E_i)}{\sqrt{(6\sqrt{2}T_2)^2 + {E_i}^2}}
\end{align}
for $i=0,2$.
The second factor contributes with:
\begin{equation} \label{eq:outer_to_R}
    U_{\text{ext},j}^\dagger(b_j^\dagger + b_j)U_{\text{ext},j} = \e{-i\omega_\text{ext}t}\ketbra{\downarrow}{\uparrow} + \e{i\omega_\text{ext}t}\ketbra{\uparrow}{\downarrow}.
\end{equation}
So:
\begin{align}
    (b_2^\dagger + b_2)(b_j^\dagger + b_j) &\rightarrow k_0 \left( \sigma_j^+ \ketbra{\tilde{0}}{\tilde{1}} + \sigma_j^- \ketbra{\tilde{1}}{\tilde{0}} + \e{-i2\omega_\text{ext}t}\sigma_j^- \ketbra{\tilde{0}}{\tilde{1}} + \e{i2\omega_\text{ext}t}\sigma_j^+ \ketbra{\tilde{1}}{\tilde{0}} \right) \\
    &\quad + k_2 \left( \sigma_j^+ \ketbra{\tilde{1}}{\tilde{2}} + \sigma_j^- \ketbra{\tilde{2}}{\tilde{1}} + \e{-i2\omega_\text{ext}t}\sigma_j^- \ketbra{\tilde{1}}{\tilde{2}} + \e{i2\omega_\text{ext}t}\sigma_j^+ \ketbra{\tilde{2}}{\tilde{1}} \right) \\
    &\simeq k_0 \left( \sigma_j^+ \ketbra{\tilde{0}}{\tilde{1}} + \sigma_j^- \ketbra{\tilde{1}}{\tilde{0}} \right) + k_2 \left( \sigma_j^+ \ketbra{\tilde{1}}{\tilde{2}} + \sigma_j^- \ketbra{\tilde{2}}{\tilde{1}} \right),
\end{align}
where we in the last equality have used the rotating wave approximation to remove the fast oscillating terms.
Since we want to look at the primed mixed states as the new tunable qutrit, we transform this to the primed basis:
\begin{align}
    (b_2^\dagger + b_2)(b_j^\dagger + b_j) &\rightarrow k_0 \sigma_j^+\braket{\tilde{1}}{1'}\ketbra{0'}{1'} + k_0\sigma_j^- \braket{1'}{\tilde{1}}\ketbra{1'}{0'} \\
    &\quad + k_2 \Big( \sigma_j^+ \braket{1'}{\tilde{1}}\braket{\tilde{2}}{1'} + \sigma_j^- \braket{1'}{\tilde{2}}\braket{\tilde{1}}{1'} \Big) \ketbra{1'} \\
    &\quad + k_2 \Big( \sigma_j^+ \braket{1'}{\tilde{1}}\braket{\tilde{2}}{2'} + \sigma_j^- \braket{1'}{\tilde{2}}\braket{\tilde{1}}{2'} \Big) \ketbra{1'}{2'} \\
    &\quad + k_2 \Big( \sigma_j^+ \braket{2'}{\tilde{1}}\braket{\tilde{2}}{1'} + \sigma_j^- \braket{2'}{\tilde{2}}\braket{\tilde{1}}{1'} \Big) \ketbra{2'}{1'} \\
    &\quad + k_2 \Big( \sigma_j^+ \braket{2'}{\tilde{1}}\braket{\tilde{2}}{2'} + \sigma_j^- \braket{2'}{\tilde{2}}\braket{\tilde{1}}{2'} \Big) \ketbra{2'} \\
    &\quad + k_0 \sigma_j^+ \braket{\tilde{1}}{2'}\ketbra{0'}{2'} + k_0 \sigma_j^- \braket{2'}{\tilde{1}}\ketbra{2'}{0'}.
\end{align}
Some of these terms look troubling, but luckily, all terms proportional with an off-diagonal overlap between the primed and tilde states will be much smaller than the diagonal ones and can be ignored. Also, all these terms are general energy non-conserving and thus could also be eliminated using a rotating wave approximation in the interaction picture. We note that the imaginary factor from the matrix elements can be eliminated by defining $ \ket{1'} \mapsto \ket{1'}_\text{new} = -i\ket{1'}_\text{old} $.
We thus end with:
\begin{align}
    (b_2^\dagger + b_2)(b_j^\dagger + b_j) &\overset{\text{R}}{\rightarrow}  \frac{k_0\Delta}{\sqrt{\Delta^2 + \left( \frac{\delta}{2} - \gamma \right)^2 }} \left( \sigma_j^+ \ketbra{0'}{1'} + \sigma_j^- \ketbra{1'}{0'} \right) \\
    &\quad + \frac{k_2\left(\frac{\delta}{2}+\gamma\right)}{2\gamma} \left( \sigma_j^+ \ketbra{1'}{2'} + \sigma_j^- \ketbra{2'}{1'} \right),
\end{align}
where the ``R'' denotes the transformation to the rotating frame.
We note that the couplings $ \ket{0'} \leftrightarrow \ket{1'} $ and $ \ket{1'} \leftrightarrow \ket{2'} $ are not equal. In fact, for small $\Delta$, the latter is a factor of $\sqrt{2}$ bigger, originating from the definition of the bosonic step operators.
\\

We can now look at the transformation of the last kind of interaction term:
\begin{align}
    \left( b_2^\dagger + b_2 \right)^2 \left( b_j^\dagger + b_j \right)^2 \rightarrow U_{\text{ext},2}^\dagger \left( b_2^\dagger + b_2\right)^2U_{\text{ext},2} \ U_{\text{ext},j}^\dagger\left(b_j^\dagger + b_j\right)^2 U_{\text{ext},j}.    
\end{align}
For the second factor, the transformation is equal to the identity for the standard qubit basis, i.e.
\begin{equation}
  U_{\text{ext},j}^\dagger(b_j^\dagger + b_j)^2U_{\text{ext},j} \mapsto 2 + \sigma_j^z.
\end{equation}
Performing the transformation for the first factor, we get:
\begin{align}
    U_{\text{ext},2}^\dagger \left( b_2^\dagger + b_2\right)^2U_{\text{ext},2} &= C_{00} \ketbra{\tilde{0}} + 3 \ketbra{\tilde{1}} + C_{22} \ketbra{\tilde{2}} \nonumber\\
    & \quad + C_{02} \Big( \e{-i2\omega_\text{ext}t} \ketbra{\tilde{0}}{\tilde{2}} + \e{i2\omega_\text{ext}t} \ketbra{\tilde{2}}{\tilde{0}} \Big).
\end{align}
Again, the last two terms are fast-rotating and can be removed.
We have defined:
\begin{equation}
	\begin{aligned}
		C_{00} &= 1 - \frac{4E_0\left(6T_2 - E_0\right)}{\left( 6\sqrt{2}T_2\right)^2 + {E_0}^2} \\
		C_{02} &= \frac{ 4E_0E_2 - 12T_2(E_0 + E_2)}{\sqrt{\left( 6\sqrt{2}T_2 \right)^2 + {E_0}^2}\sqrt{\left( 6\sqrt{2}T_2 \right)^2 {E_2}^2}}\\
		C_{22} &= 1 - \frac{4E_2\left(6T_2 - E_2\right)}{\left( 6\sqrt{2}T_2\right)^2 + {E_2}^2} .
	\end{aligned}
\end{equation}
Moving to the primed basis, we find:
\begin{align*}
    \left( b_2^\dagger + b_2\right)^2 &\rightarrow C_{00} \ketbra{0'} + \Big( 3 \abs{\braket{1'}{\tilde{1}}}^2 + C_{22}\abs{\braket{1'}{\tilde{2}}}^2 \Big) \ketbra{1'} \nonumber\\
    &\quad + \Big( 3 \braket{1'}{\tilde{1}}\braket{\tilde{1}}{2'} + C_{22} \braket{1'}{\tilde{2}}\braket{\tilde{2}}{2'} \Big) \ketbra{1'}{2'} + \Big( 3 \braket{2'}{\tilde{1}}\braket{\tilde{1}}{1'} + C_{22} \braket{2'}{\tilde{2}}\braket{\tilde{2}}{1'} \Big) \ketbra{2'}{1'} \nonumber\\
    &\quad + \Big( 3\abs{\braket{2'}{\tilde{1}}}^2 + C_{22} \abs{\braket{2'}{\tilde{2}}}^2 \Big) \ketbra{2'}.
\end{align*}
Again, we will ignore the energy non-conserving terms proportional to an off-diagonal overlap. We will, though, keep the terms proportional to a non-diagonal overlap in the energy conserving terms for accuracy of the final Hamiltonian.
Evaluating the matrix elements gives:
\begin{align}
    \left( b_2^\dagger + b_2\right)^2 &\overset{\text{R}}{\rightarrow} C_{00} \ketbra{0'} + \frac{3\Delta^2 + C_{22} \left( -\frac{\delta}{2} + \gamma \right)^2}{\Delta^2 + \left( -\frac{\delta}{2} + \gamma \right)^2} \ketbra{1'} \nonumber\\
    &\quad + \frac{3\Delta^2 + C_{22} \left( \frac{\delta}{2} + \gamma \right)^2}{\Delta^2 + \left( \frac{\delta}{2} + \gamma \right)^2} \ketbra{2'}.
\end{align}
We are now ready to look at the full transformed Hamiltonian.

\subsection{Full Hamiltonian}

We can now write down the full Hamiltonian for the system when coupled to an external field mixing the first and second excited state in the rotating frame of the Hamiltonian $H_\text{ext} = -\frac{3\omega_\text{ext}}{2 } \ketbra{\tilde{0}} -\frac{\omega_\text{ext}}{2} \ketbra{\tilde{1}} + \frac{\omega_\text{ext}}{2} \ketbra{\tilde{2}} $. We start from the Hamiltonian in equation \eqref{eq:H_b} and now insert how the factors containing the bosonic step operators transform under such a transformation, as calculated in the previous section. To sum, up, we found:
\begin{equation}
	\begin{aligned} \label{eq:qutrit_mappings}
		T_1b_2^\dagger b_2 - T_2\left( b_2^\dagger + b_2 \right)^4 + iT_\text{ext} \cos(\omega_\text{ext} t)\left( b_2^\dagger - b_2 \right) &\mapsto (E_1' - E_0')\ketbra{1'} + (E_2' - E_0')\ketbra{2'} \\
     \sqrt{\frac{4\alpha_j}{C_j}}\left( b_j^\dagger b_j + \frac{1}{2} \right) - \frac{\beta_j}{4\alpha_jC_j}\left( b_j^\dagger + b_j \right)^4 &\mapsto \frac{1}{2}\left( \sqrt{\frac{4\alpha_j}{C_j}} - \frac{3\beta_j}{\alpha_jC_j} - \omega_\text{ext} \right) \sigma_j^z\\
		(b_2^\dagger + b_2)(b_j^\dagger + b_j) &\mapsto  \frac{k_0\Delta}{\sqrt{\Delta^2 + \left( \frac{\delta}{2} - \gamma \right)^2 }} \left( \sigma_j^+ \ketbra{0'}{1'} + \sigma_j^- \ketbra{1'}{0'} \right) \\
    &\phantom{\mapsto} + \frac{k_2\left(\frac{\delta}{2}+\gamma\right)}{2\gamma} \left( \sigma_j^+ \ketbra{1'}{2'} + \sigma_j^- \ketbra{2'}{1'} \right)\\
		\left( b_2^\dagger + b_2 \right)^2\left( b_j^\dagger + b_j \right)^2 &\mapsto \Big( C_{00}\mathbb{I}_2 + D_1\ketbra{1'} + D_2\ketbra{2'} \Big)\Big( 2 + \sigma_j^z \Big) ,
	\end{aligned}
\end{equation}
where
\begin{equation} \label{eq:D_constants}
	\begin{split}
    D_1 &=  \frac{3\Delta^2 + C_{22} \left( -\frac{\delta}{2} + \gamma \right)^2}{\Delta^2 + \left( -\frac{\delta}{2} + \gamma \right)^2} - C_{00} \\
    D_2 &=  \frac{3\Delta^2 + C_{22} \left( \frac{\delta}{2} + \gamma \right)^2}{\Delta^2 + \left( \frac{\delta}{2} + \gamma \right)^2} - C_{00}.
\end{split}
\end{equation}
We remind the reader that we have defined the excited state as $ \ket{\uparrow} $ and the ground state as $ \ket{\downarrow} $, which explains the sign in front of the $ \sigma^z $s.
\noindent We can now write out the full Hamiltonian in the rotating frame, changing the indices from ${1,2,3}$ to $ {L,M,R} $ for the sake of visualizing the system as a chain of two qubits with a qutrit in the middle:
\begin{equation} \label{eq:H_full}
	\begin{aligned}
    H_\text{full} &= \frac{\Delta_L}{2} \sigma_L^z + \Delta_M \ketbra{1'} + (\Delta_M + \delta_M)\ketbra{2'} + \frac{\Delta_R}{2} \sigma_R^z \\
    &\hspace{1cm} + J_{LM_{01}}\left( \sigma_L^-\ketbra{1'}{0'} + \sigma_L^+\ketbra{0'}{1'} \right) +J_{RM_{01}}\left( \sigma_R^-\ketbra{1'}{0'} + \sigma_R^+\ketbra{0'}{1'} \right) \\
    &\hspace{1cm}+ J_{LM_{12}}\left( \sigma_L^-\ketbra{2'}{1'} + \sigma_L^+\ketbra{1'}{2'} \right)+ J_{RM_{12}}\left( \sigma_R^-\ketbra{2'}{1'} + \sigma_R^+\ketbra{1'}{2'} \right) \\
    &\hspace{1cm}+ J_{LM}^{(z)}\sigma_L^z\left( D_1\ketbra{1'} + D_2\ketbra{2'} \right) + J_{RM}^{(z)}\sigma_R^z\left( D_1\ketbra{1'} + D_2\ketbra{2'} \right),
\end{aligned}
\end{equation}
Here, the diagonal constants are
\begin{equation}
  \begin{aligned}\label{eq:diagonal_constants}
    \Delta_L &= -\frac{3\beta_1}{\alpha_1C_1} + \sqrt{\frac{4\alpha_1}{C_1}} - \frac{C_{00}}{64}\frac{E_{J1}\cos(\frac{\Phi_{\Sigma 1}}{2})}{\sqrt{\alpha_1\alpha_2C_1(C_1+C_2+C_\text{ext})}} - \omega_\text{ext} \\
    \Delta_M &= E_1'-E_0' - \frac{D_1}{64}\frac{E_{J1}\cos(\frac{\Phi_{\Sigma 1}}{2})}{\sqrt{\alpha_1\alpha_2C_1(C_1+C_2+C_\text{ext})}} - \frac{D_1}{64}\frac{E_{J2}\cos(\frac{\Phi_{\Sigma 2}}{2})}{\sqrt{\alpha_3\alpha_2C_2(C_1+C_2+C_\text{ext})}} \\
    \delta_M &= 2\gamma - \frac{D_2-D_1}{64}\frac{E_{J1}\cos(\frac{\Phi_{\Sigma 1}}{2})}{\sqrt{\alpha_1\alpha_2C_1(C_1+C_2+C_\text{ext})}} - \frac{D_2-D_1}{64}\frac{E_{J2}\cos(\frac{\Phi_{\Sigma 2}}{2})}{\sqrt{\alpha_3\alpha_2C_2(C_1+C_2+C_\text{ext})}} \\
    \Delta_R &= - \frac{3\beta_3}{\alpha_3C_2} + \sqrt{\frac{4\alpha_3}{C_2}} - \frac{C_{00}}{64}\frac{E_{J2}\cos(\frac{\Phi_{\Sigma 2}}{2})}{\sqrt{\alpha_3\alpha_2C_2(C_1+C_2+C_\text{ext})}} - \omega_\text{ext} ,
  \end{aligned}
\end{equation}
while the $D_i$ constants are defined in \eqref{eq:D_constants} and the rest are

\begin{equation}\label{eq:interaction_constants}
	\begin{aligned}
		J_{LM_{01}} &= \left( \frac{1}{8L_1} - \frac{1}{8\tilde{L}_1} \right)\frac{k_0}{2(\alpha_1\alpha_2C_1(C_1+C_2+C_\text{ext}))^{\frac{1}{4}}}\frac{\Delta}{\sqrt{\Delta^2 + \left( -\frac{\delta}{2}+\gamma \right)^2 }} \\
		J_{RM_{01}} &= \left( \frac{1}{8L_2} - \frac{1}{8\tilde{L}_2} \right)\frac{k_0}{2(\alpha_3\alpha_2C_2(C_1+C_2+C_\text{ext}))^{\frac{1}{4}}}\frac{\Delta}{\sqrt{\Delta^2 + \left( -\frac{\delta}{2}+\gamma \right)^2 }} \\
		J_{LM_{12}} &= \left( \frac{1}{8L_1} - \frac{1}{8\tilde{L}_1} \right)\frac{k_2}{4(\alpha_1\alpha_2C_1(C_1+C_2+C_\text{ext}))^{\frac{1}{4}}}\frac{ \frac{\delta}{2} + \gamma}{\gamma} \\
		J_{RM_{12}} &= \left( \frac{1}{8L_2} - \frac{1}{8\tilde{L}_2} \right)\frac{k_2}{4(\alpha_3\alpha_2C_2(C_1+C_2+C_\text{ext}))^{\frac{1}{4}}}\frac{ \frac{\delta}{2} + \gamma}{\gamma} \\
		J_{LM}^{(z)} &= -\frac{E_{J1}}{128}\frac{\cos(\frac{\Phi_{\Sigma 1}}{2})}{\sqrt{\alpha_1\alpha_2C_1(C_1+C_2+C_\text{ext})}} \\
		J_{RM}^{(z)} &= -\frac{E_{J2}}{128}\frac{\cos(\frac{\Phi_{\Sigma 2}}{2})}{\sqrt{\alpha_3\alpha_2C_2(C_1+C_2+C_\text{ext})}}.
	\end{aligned}
\end{equation}

To sum up, we have now, starting from the circuit in Supplementary Fig. \ref{fig:big_circuit}, calculated the resulting Hamiltonian and added a dynamical tuning of the qutrit via driving of the first and second excited bare qutrit energy levels. In doing so, we first transformed to the rotating picture with respect to the driving field and then to the new mixed qutrit eigenstates, finally obtaining the Hamiltonian above.

\section*{Supplementary Note 2: Circuit parameters used in the simulations}

In Table \ref{tbl:realistic_parameters}, we have included a list of realistic circuit parameters and the corresponding spin-model parameters they yield. We can split this into three parts, a circuit with Hamiltonian parameters suited to implementing the dissociation procedure, a circuit implementing the \textsc{acswap} gate, and a circuit with all levels off-resonant, suitable to implementing the STIRAP procedure and \textsc{ccz} and holonomic gates. Note that for the dissociation and \textsc{acswap} procedures, we are working in the frame rotating with the AC Stark drive frequency, $\omega_\text{ext}$, and thus the energy terms are reduced accordingly. This of course does not change the (non-trivial) dynamics, since only the relative energy differences are important.

\begin{table}[h!]
\centering
    \begin{tabular}{| p{3.8cm} | p{6.5cm} | p{6.5cm} | }
        \hline
         & Circuit parameters & Spin-model parameters \\ \hline
        Dissociation, first part (STIRAP)
        &\vspace{-0.5cm}\begin{flushleft}
            $L_{1,2} = 19.999\:\si{\nano\henry},\ \linebreak \tilde{L}_{1,2} = 18.713\:\si{\nano\henry},\ \linebreak E_{J1,2}/(2\pi) = 185.91\:\si{\giga\hertz},\ \linebreak E_{Jq1,3}/(2\pi) = 79.515\:\si{\giga\hertz},\ \linebreak E_{Jq2}/(2\pi) = 27.441\:\si{\giga\hertz},\ \linebreak C_{1,2} = 55.816\:\si{\femto\farad},\ \linebreak\Phi_{\Sigma 1,2} = -0.43452\:\Phi_0,\ \linebreak C_\text{ext} = 2.3411\:\si{\femto\farad},\ \linebreak \text{Dynamical driving off}$
        \end{flushleft} 
        &  \vspace{-0.5cm}\begin{flushleft}
            $\Delta_L/(2\pi) = 15.271\:\si{\giga\hertz},\ \linebreak \Delta_M/(2\pi) = 13.841\:\si{\giga\hertz},\ \linebreak \delta_M /(2\pi) = 13.671\:\si{\giga\hertz},\ \linebreak \Delta_R/(2\pi) = 15.271\:\si{\giga\hertz},\ \linebreak J_{\alpha M_{01}}/(2\pi) = 9.2737\:\si{\mega\hertz},\ \linebreak J_{\alpha M_{12}}/(2\pi) = 12.996\:\si{\mega\hertz},\ \linebreak J_{\alpha M}^{(z)}/(2\pi) = -20.433\:\si{\mega\hertz},\ \linebreak D_1 = 1.9877,\ \linebreak D_2 = 3.9754,\ \linebreak \max(\Omega_{1,2})/(2\pi) = 20\:\si{\mega\hertz}$
        \end{flushleft}
        \\ \hline
        Dissociation, second part (two-photon resonance)
        &\vspace{-0.5cm}\begin{flushleft}
            $L_{1,2} = 20.000\:\si{\nano\henry},\ \linebreak \tilde{L}_{1,2} = 16.949\:\si{\nano\henry},\ \linebreak E_{J1,2}/(2\pi) = 32.371\:\si{\giga\hertz},\ \linebreak E_{Jq1,3}/(2\pi) = 109.74\:\si{\giga\hertz},\ \linebreak E_{Jq2}/(2\pi) = 108.94\:\si{\giga\hertz},\ \linebreak C_{1,2} = 55.935\:\si{\femto\farad},\ \linebreak\Phi_{\Sigma 1,2} = 0.27790\:\Phi_0,\ \linebreak C_\text{ext} = 71.773\:\si{\femto\farad},\ \linebreak
            A_\text{ext}/(4\alpha_2 (C_1+C_2+C_\text{ext}))^{-1/4} = 0.1000,\ \linebreak \omega_{\text{ext}}/(2\pi) = 14.674\:\si{\giga\hertz}$
        \end{flushleft} 
        &  \vspace{-0.5cm}\begin{flushleft}
            In the frame rotating with $\omega_\text{ext}$: \linebreak 
            $\Delta_L/(2\pi) = 0.46529\:\si{\giga\hertz},\ \linebreak \Delta_M/(2\pi) = 0.088376\:\si{\giga\hertz},\ \linebreak \delta_M /(2\pi) = 0.84222\:\si{\giga\hertz},\ \linebreak \Delta_R/(2\pi) = 0.46529\:\si{\giga\hertz},\ \linebreak J_{\alpha M_{01}}/(2\pi) = 15.0203\:\si{\mega\hertz},\ \linebreak J_{\alpha M_{12}}/(2\pi) = 17.114\:\si{\mega\hertz},\ \linebreak J_{\alpha M}^{(z)}/(2\pi) = -6.4890\:\si{\mega\hertz},\ \linebreak D_1 = 2.6636,\ \linebreak D_2 = 3.3015$
        \end{flushleft}
        \\ \hline
        \textsc{ccnot}/\textsc{ccz}
        &\vspace{-0.5cm}\begin{flushleft}
            Same as during STIRAP.
        \end{flushleft} 
        &  \vspace{-0.5cm}\begin{flushleft}
            Same as for STIRAP, except \linebreak $\Omega_1/(2\pi) = 6\:\si{\mega\hertz}$
        \end{flushleft}
        \\ \hline
        \textsc{acswap} \newline(first part of \textsc{ccswap})
        & \vspace{-0.5cm}\begin{flushleft}
            $L_{1,2} = 16.049\:\si{\nano\henry},\ \linebreak \tilde{L}_{1,2} = 18.988\:\si{\nano\henry},\ \linebreak E_{J1,2}/(2\pi) = 56.08\:\si{\giga\hertz},\ \linebreak E_{Jq1,3}/(2\pi) = 155.22\:\si{\giga\hertz},\ \linebreak E_{Jq2}/(2\pi) = 135.33\:\si{\giga\hertz},\ \linebreak C_{1,2} = 56.619\:\si{\femto\farad},\ \linebreak \Phi_{\Sigma 1,2} = 0.49999\:\Phi_0,\ \linebreak C_\text{ext} = 90.011\:\si{\femto\farad},\ \linebreak A_\text{ext}/(4\alpha_2 (C_1+C_2+C_\text{ext}))^{-1/4} = 0.10000,\ \linebreak \omega_{\text{ext}}/(2\pi) = 14.367\:\si{\giga\hertz}$
        \end{flushleft}
        & \vspace{-0.5cm}\begin{flushleft}
            In the frame rotating with $\omega_\text{ext}$: \linebreak
            $\Delta_L/(2\pi) = 0.9113\:\si{\giga\hertz},\ \linebreak \Delta_M/(2\pi) = -0.0798\:\si{\giga\hertz},\ \linebreak \delta_M /(2\pi) = 0.9121\:\si{\giga\hertz},\ \linebreak \Delta_R/(2\pi) = 0.9113\:\si{\giga\hertz},\ \linebreak J_{\alpha M_{01}}/(2\pi) = 14.311\:\si{\mega\hertz},\ \linebreak J_{\alpha M_{12}}/(2\pi) = 15.173\:\si{\mega\hertz},\ \linebreak J_{\alpha M}^{(z)}/(2\pi) = -0.601\:\si{\kilo\hertz},\ \linebreak D_1 = 2.8377,\ \linebreak D_2 = 3.12544$
        \end{flushleft}
        \\ \hline
        Holonomic gate \phantom{abc} \linebreak (double-controlled) & \vspace{-0.5cm}\begin{flushleft}
            \vspace{0.0pt}  Same as during STIRAP
        \end{flushleft}
        & \vspace{-0.5cm}\begin{flushleft}
            Same as for STIRAP, except \linebreak $\Omega_{1,2}/(2\pi) = 15\:\si{\mega\hertz}$
        \end{flushleft} \\ \hline

    \end{tabular}
    \caption{A table of realistic parameters and the corresponding Hamiltonian parameters used in each implementation. In all simulations, we have included finite relaxation and coherence times set to $T_1 = \SI{31}{\micro\second}$ and $T_2 = \SI{35}{\micro\second}$, respectively. The parameters of the second part of the \textsc{cswap} is not shown, but is obtained by varying only the external parameters relative to the first part. The resulting parameters are similar to what is shown for the \textsc{ccnot} gate implementation.}
    \label{tbl:realistic_parameters}
\end{table}

\section*{Supplementary Note 3: Direct Dissociation to Entanglement}

We wish to obtain an entangled state between the state where qubit one and three is in the excited state and the state where they are both in the ground state. Since the Hamiltonian conserves the total projection of spin, we start in the state $\ket{\downarrow 2 \downarrow}$. We look at the matrix representation of our Hamiltonian in the basis $ \left\{\, \ket{\downarrow 2 \downarrow}, \ket{\downarrow 1 \uparrow}, \ket{\uparrow 1 \downarrow}, \ket{\uparrow 0 \uparrow}\, \right\} $. If we assume the system to be symmetric, i.e. we do not distinguish between $\alpha=L$ and $\alpha=R$, and further assume the states to be resonant, i.e. $\Delta_M = \Delta_R = \Delta_L = \delta_M$, the contribution from these energy terms is proportional to the identity and we are left with the matrix representation of the XX and ZZ-terms, which we can easily diagonalize to find the eigenvalues. In the simplified case $J_{\alpha M_{01}} = J_{\alpha M_{12}} = J_{\alpha M} $ and $ D_1 = D_2 = 1 $ (This is not essential, and is only chosen as to ease the readability of the analysis. The equation for $D_2J_{\alpha M}^{(z)}$ in the main text is the result in the non-simplified case), the eigenvalues are:
\begin{align}
  E_1 = 0 \quad , \quad E_2 = -2 J_{\alpha M}^{(z)} \quad , \quad E_3 = -J_{\alpha M}^{(z)} - \lambda \quad \text{and} \quad E_4 = -J_{\alpha M}^{(z)} + \lambda,
\end{align}
where
\begin{equation}\label{eq:lambda}
  \lambda = \sqrt{4{J_{\alpha M}}^2 + {J_{\alpha M}^{(z)}}^2}.
\end{equation}
The associated eigenvectors are
\begingroup
\renewcommand*{\arraystretch}{1.5}
\begin{align}
  \ket{v_1} = \frac{1}{\sqrt{2}}
  \begin{pmatrix} 0 \\ 0 \\ -1 \\ 1 \end{pmatrix}
  \quad , \quad \ket{v_2} &= \frac{1}{\sqrt{2}}
  \begin{pmatrix} -1 \\ 1 \\ 0 \\ 0 \end{pmatrix}
  \quad , \quad \ket{v_3} = \sqrt{\frac{\lambda -J_{\alpha M}^{(z)}}{4\lambda}}
  \begin{pmatrix} \frac{2 J_{\alpha M}}{J_{\alpha M}^{(z)} - \lambda} \\ \frac{2 J_{\alpha M}}{J_{\alpha M}^{(z)} - \lambda} \\ 1 \\ 1 \end{pmatrix} \nonumber \\
  \quad \text{and} \quad \ket{v_4} &= \sqrt{\frac{\lambda +J_{\alpha M}^{(z)}}{4\lambda}}
  \begin{pmatrix} \frac{2 J_{\alpha M}}{J_{\alpha M}^{(z)} + \lambda} \\ \frac{2 J_{\alpha M}}{J_{\alpha M}^{(z)} + \lambda} \\ 1 \\ 1\end{pmatrix}.
\end{align}
\endgroup
We can now find
\begin{align}
    \braket{f}{\text{e}^{-iHt} | \downarrow 2 \downarrow} &= \sum_{n,m=1}^4 \braket{f}{v_n}\braket{v_n | \text{e}^{-iHt}}{v_m} \braket{v_m}{\downarrow 2 \downarrow} \nonumber \\
    &=\sum_{n=1}^4 \braket{f}{v_n} \text{e}^{-iE_nt}\braket{v_n}{\downarrow 2 \downarrow},
\end{align}
where $\ket{f}$ is one of the four basis states mentioned earlier. From this, the complete time development of $\ket{\downarrow 2 \downarrow}$ is recovered analytically. Specifically, we have
\begin{align*}
    \abs{\braket{\downarrow 2 \downarrow}{\text{e}^{-iHt} | \downarrow 2 \downarrow}}^2 =  \frac{1}{4}\Big( \cos(J_{\alpha M}^{(z)}t) + \cos(\lambda t) \Big)^2 + \frac{1}{4}\Big( \sin(J_{\alpha M}^{(z)}t) + \frac{J_{\alpha M}^{(z)}}{\lambda}\sin(\lambda t) \Big)^2
\end{align*}
and
\begin{align*}
\abs{\braket{\uparrow 0 \uparrow}{\text{e}^{-iHt} | \downarrow 2 \downarrow}}^2 =  \frac{1}{4}\Big( \cos(J_{\alpha M}^{(z)}t) - \cos(\lambda t) \Big)^2 + \frac{1}{4}\Big( \sin(J_{\alpha M}^{(z)}t) - \frac{J_{\alpha M}^{(z)}}{\lambda}\sin(\lambda t) \Big)^2.
\end{align*}

Because $\lambda = \sqrt{4{J_{\alpha M}}^2 + {J_{\alpha M}^{(z)}}^2} $ we, for a general $J_{\alpha M}^{(z)} $, expect the probabilities to have a chaotic behavior. We ask ourselves if there exists a certain value of $J_{\alpha M}^{(z)}>0 $ where we can find a time $t$ at which
\begin{align} \label{eq:condition_1}
  &\abs{\braket{\downarrow 2 \downarrow}{\text{e}^{-iHt} | \downarrow 2 \downarrow}}^2 =
    \abs{\braket{\uparrow 0 \uparrow}{\text{e}^{-iHt} | \downarrow 2 \downarrow}}^2
    =\frac{1}{2}
\end{align}
and
\begin{align} \label{eq:condition_2}
    \frac{\text{d}}{\text{d}t}\abs{\braket{\downarrow 2 \downarrow}{\text{e}^{-iHt} | \downarrow 2 \downarrow}}^2 =
    \frac{\text{d}}{\text{d}t}\abs{\braket{\uparrow 0 \uparrow}{\text{e}^{-iHt} | \downarrow 2 \downarrow}}^2
    =0.
\end{align}
From the first equality in \eqref{eq:condition_1}, we find the condition
\begin{align}\label{eq:con_1_2}
    \cos(\lambda t)\cos(J_{\alpha M}^{(z)}t) = - \frac{J_{\alpha M}^{(z)}}{\lambda}\sin(\lambda t)\sin(J_{\alpha M}^{(z)}t),
\end{align}
while we from equation \eqref{eq:condition_2} get
\begin{align}\label{eq:con_2_2}
    &\sin(\lambda t)\left( \cos(J_{\alpha M}^{(z)}t) + \cos(\lambda t) \right) = 0 = \sin(\lambda t)\left( \cos(J_{\alpha M}^{(z)}t) - \cos(\lambda t) \right) \nonumber \\
    &\Rightarrow \quad\sin(\lambda t) = 0 \quad \quad \text{or} \quad \quad \cos(J_{\alpha M}^{(z)}t)=\cos(\lambda t) = 0.
\end{align}
Here, the second option implies that $\sin(J_{\alpha M}^{(z)}t) = \pm \sin(\lambda t) = \pm 1 \neq 0$, which is not compatible with the condition \eqref{eq:con_1_2}. Therefore, we must have $ \sin(\lambda t) = 0 $ and from \eqref{eq:con_1_2}  also $\cos(J_{\alpha M}^{(z)}t) = 0$. This means that we have two conditions on our time $t>0$:
\begin{align}
    t=n\frac{\pi}{\lambda} \quad \text{and} \quad t=m\frac{\pi}{2J_{\alpha M}^{(z)}} \ ,\ n\in \mathbb{N}\,,\ m=1,3,5,\dots
\end{align}
We solve this for $J_{\alpha M}^{(z)}$:
\begin{align}
    n\frac{\pi}{\lambda} &= m\frac{\pi}{2J_{\alpha M}^{(z)}} \Rightarrow 4n^2{J_{\alpha M}^{(z)}}^2 = (4{J_{\alpha M}}^2 + {J_{\alpha M}^{(z)}}^2)m^2\nonumber
    \\ &\Rightarrow J_{\alpha M}^{(z)} = \frac{2J_{\alpha M}}{\sqrt{4 n^2 / m^2 - 1}},\label{eq:J_ZZ}
\end{align}
where $n=1,2,3,\dots$ and $m = 1, 3, 5,\dots$  with the further condition to keep the time real $4 n^2 / m^2 > 1 \Rightarrow n > m/2 $.
This is plotted for $n=m=1$ in Supplementary Fig. \ref{fig:n1m1} for arbitrary values of parameters for proof of concept.
\begin{figure}[!htbp]
  \centering
  \includegraphics[width=0.75\textwidth]{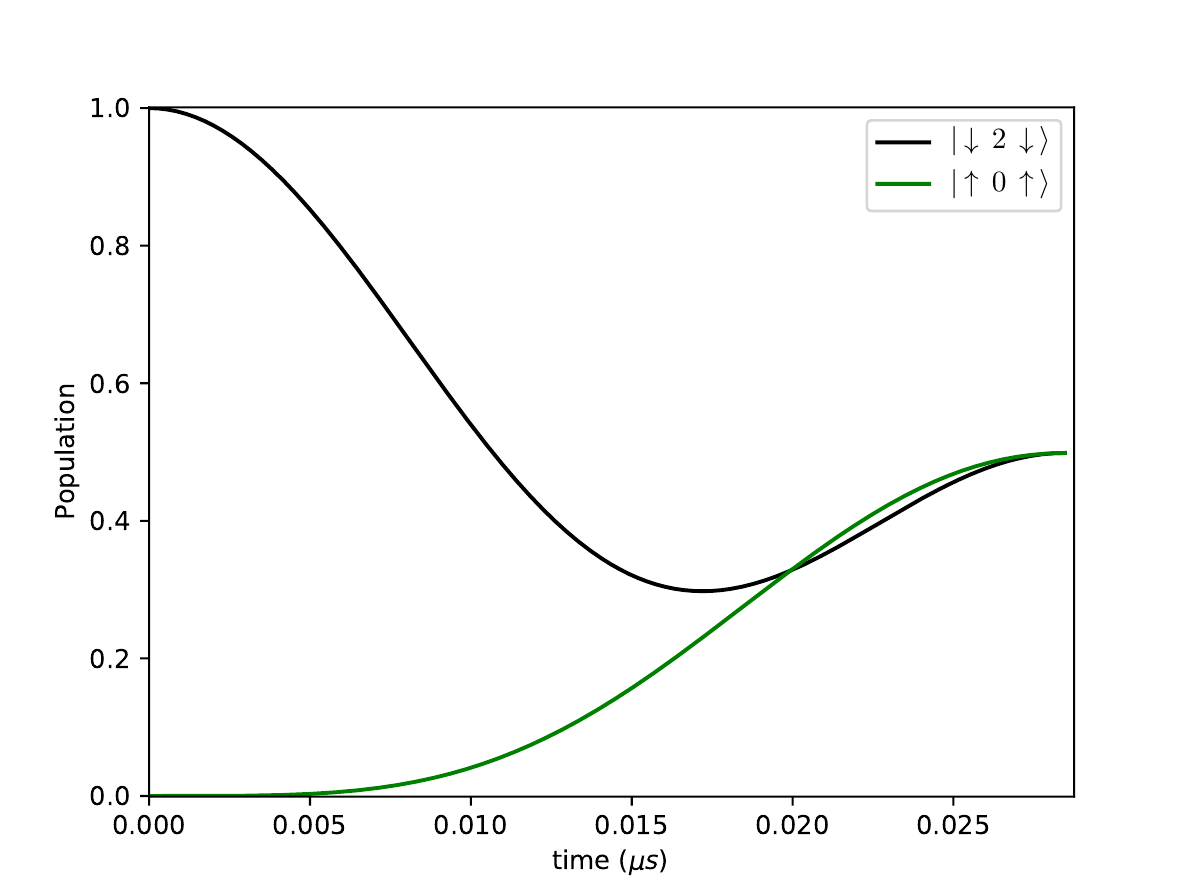}
  \caption{Probability of occupation in the case $J_{\alpha M}^{(z)} = 2J_{\alpha M}/\sqrt{3} $. We notice that the desired entangled state is reached at $t = \pi /\lambda = \pi /2J_{\alpha M}^{(z)} \approx 1.36 \frac{1}{J_\alpha M} \approx \SI{0.028}{\micro\second} $.}
  \label{fig:n1m1}
\end{figure}

\section*{Supplementary Note 4: Analysis of the CSWAP}\label{sec:appendix_CSWAP}

Assuming that the excited states of the outer qubits are in resonance with the second excited state of the qutrit ($\Delta_L \simeq \delta_M \simeq \Delta_R $) while the qutrit ground state is far detuned from the rest, we can move to the interaction picture of the diagonal in equation \eqref{eq:H_full}, ignoring these terms and also all terms involving the ground state ($\ket{0}$) of the qutrit. We also choose to assume symmetry between both ends of the chain, so e.g. $ J_{L M_{12}} = J_{R M_{12}} = J_{\alpha M_{12}} $, and that $J_{\alpha M}^{(z)} = 0$. A non-zero ZZ-coupling will make the states with the qubits in the same state slightly detuned, introducing a non-perfect transition. This can be remedied a bit by choosing $\Delta_\alpha = \delta_M - 2\frac{D_1+D_2}{2}J_{\alpha M}^{(z)}$, but because in general $D_1 \neq D_2$, this detuning can not be fixed completely. In the basis $\{\ket{\downarrow 1 \downarrow},\ket{\uparrow 1 \downarrow}, \ket{\downarrow 2 \downarrow}, \ket{\downarrow 1 \uparrow}, \ket{\uparrow 2 \downarrow}, \ket{\uparrow 1 \uparrow},\ket{\downarrow 2 \uparrow},\ket{\uparrow 2 \uparrow}\}$, the interaction part of the Hamiltonian can be written very simply:
\begin{equation}
	H_I = \begin{pmatrix}[c | ccc | ccc | c]
		0 & 0 & 0 & 0 & 0 & 0 & 0 & 0 \\ \hline
		0 & 0 & J_{\alpha M_{12}} & 0 & 0 & 0 & 0 & 0 \\
		0 & J_{\alpha M_{12}} & 0 & J_{\alpha M_{12}} & 0 & 0 & 0 & 0 \\
		0 & 0 & J_{\alpha M_{12}} & 0 & 0 & 0 & 0 & 0 \\ \hline
		0 & 0 & 0 & 0 & 0 & J_{\alpha M_{12}} & 0 & 0 \\
		0 & 0 & 0 & 0 & J_{\alpha M_{12}} & 0 & J_{\alpha M_{12}} & 0 \\
		0 & 0 & 0 & 0 & 0 & J_{\alpha M_{12}} & 0 & 0 \\ \hline
		0 & 0 & 0 & 0 & 0 & 0 & 0 & 0
	\end{pmatrix} 	.
\end{equation}
Notice the block matrix form where the two $3\times3$ matrices represent irreducible subspaces. Furthermore, they are equal in form, so when performing the time evolution $\e{-iH_I t} $ we only have to exponentiate one of these irreducible sub-matrices.
It is easily found that:
\begin{equation}
		\exp[-i \begin{pmatrix}
			0 & J_{\alpha M_{12}} & 0 \\
			J_{\alpha M_{12}} & 0 & J_{\alpha M_{12}} \\
			0 & J_{\alpha M_{12}} & 0
		\end{pmatrix}t] = \frac{1}{2} \begin{pmatrix}
			\cos(\sqrt{2}J_{\alpha M_{12}} t) + 1 & -i\sqrt{2}\sin(\sqrt{2}J_{\alpha M_{12}} t) & \cos(\sqrt{2}J_{\alpha M_{12}} t) - 1 \\
			-i\sqrt{2}\sin(\sqrt{2}J_{\alpha M_{12}} t) & 2\cos(\sqrt{2}J_{\alpha M_{12}} t) & -i\sqrt{2}\sin(\sqrt{2}J_{\alpha M_{12}} t)\\
			\cos(\sqrt{2}J_{\alpha M_{12}} t) - 1 & -i\sqrt{2}\sin(\sqrt{2}J_{\alpha M_{12}} t) & \cos(\sqrt{2}J_{\alpha M_{12}} t) + 1
		\end{pmatrix}.
\end{equation}
Letting the qutrit start in the ground state, we can now find
\begin{align}
	\e{-iH_I t} \ket{\downarrow 1 \downarrow} &= \ket{\downarrow 1 \downarrow} \\
	\e{-iH_I t} \ket{\uparrow 1 \downarrow} &= \cos[2](\frac{J_{\alpha M_{12}}}{\sqrt{2}}t)\ket{\uparrow 1 \downarrow} - \frac{i}{\sqrt{2}}\sin(\sqrt{2}J_{\alpha M_{12}}t)\ket{\downarrow 2 \downarrow} - \sin[2](\frac{J_{\alpha M_{12}}}{\sqrt{2}}t)\ket{\downarrow 1 \uparrow} \\
	\e{-iH_I t} \ket{\uparrow 1 \uparrow} &= -\frac{i}{\sqrt{2}}\sin(\sqrt{2}J_{\alpha M_{12}}t)\Big( \ket{\uparrow 2 \downarrow} + \ket{\downarrow 2 \uparrow} \Big) + \cos(\sqrt{2}J_{\alpha M_{12}}t)\ket{\uparrow 1 \uparrow}.
\end{align}
We wish to obtain a \mbox{\textsc{-cswap}}, so we require $ \abs{\matrixel{\downarrow 1 \uparrow}{\e{-iH_It}}{\uparrow 1 \downarrow}}^2 = 1 $, thus we must choose the operation time to be $T = \frac{\pi}{\sqrt{2}J_{\alpha M_{12}}} $. Inserting this into the equations above, we get
\begin{align}
	\matrixel{\downarrow 1 \uparrow}{\e{-iH_IT}}{\uparrow 1 \downarrow} = \matrixel{\uparrow 1 \uparrow}{\e{-iH_IT}}{\uparrow 1 \uparrow} = -1 = - \matrixel{\downarrow 1 \downarrow}{\e{-iH_IT}}{\downarrow 1 \downarrow}.
\end{align}
This last matrix element has the wrong sign, so if we tried to perform a \mbox{\textsc{-cswap}} between the outer qubits in different superpositions of the `up' and `down' state, we would obtain an unfortunate relative sign change.
This can be seen by assuming the left qubit starts in the superposition
\begin{equation}
	\ket{\psi_\text{L}} = a\ket{\uparrow} + b \ket{\downarrow},
\end{equation}
while the right qubit starts in the superposition
\begin{equation}
	\ket{\psi_\text{R}} = c\ket{\uparrow} + d \ket{\downarrow}.
\end{equation}
Here, $a,b,c$ and $d$ are complex coefficients. The total starting state of the system is then the factorizable state
\begin{equation}
	\left( a\ket{\uparrow} + b\ket{\downarrow} \right)\ket{1}\left( c\ket{\uparrow} + d\ket{\downarrow} \right) = ac\ket{\uparrow1\uparrow} + ad\ket{\uparrow1\downarrow} + bc\ket{\downarrow1\uparrow} + bd\ket{\downarrow1\downarrow}.
\end{equation}
Operating on this state with the unitary operator $ \e{-iH_I T} $ now gives:
\begin{equation}
	- ac\ket{\uparrow 1 \uparrow} - ad\ket{\downarrow1\uparrow} - bc\ket{\uparrow1\downarrow} + bd\ket{\downarrow 1 \downarrow}.
\end{equation}
This state is not factorizable because of the sign difference between the first term and the rest, and therefore does not have a simple interpretation as the system where a \textsc{swap} operation has been performed.

This can be fixed by first operating with a \textsc{ccz} gate on the system so that only this specific state ($ \ket{\downarrow 1 \downarrow} $) obtains a sign change. This of course requires that the states are first moved out of resonance, but this is already required in order to catch the system in the swapped state. We then obtain the state
\begin{equation}
	- ac\ket{\uparrow 1 \uparrow} - ad\ket{\downarrow 1 \uparrow} - bc\ket{\uparrow 1 \downarrow} - bd\ket{\downarrow 1 \downarrow} = -\left( c\ket{\uparrow} + d\ket{\downarrow} \right)\ket{1}\left( a\ket{\uparrow} + b\ket{\downarrow} \right).
\end{equation}
It is clear that the states of the outer qubits have been swapped, just as we wanted!

We thus end up with a \textsc{cswap} gate in the interaction picture of $H_0$ with a two-step operation scheme and with the control ``bit'' being comprised of the states $\ket{0}$ and $\ket{1}$ of the qutrit (or, equivalently, the states $\ket{1}$ and $\ket{2}$ if we instead apply the \textsc{ccz} so only the state $\ket{\uparrow 1 \uparrow}$ receives a sign change). By symmetry, the topmost excited qutrit state $\ket{2}$ will also allow transfer, but here the state $\ket{\uparrow 2 \uparrow}$ will need a sign change.

A simulation of the \textsc{cswap}/Fredkin gate is shown in the main text, where we get a fidelity of around $0.95$ for a perfect \textsc{swap} operation when the qutrit starts in the first excited state, as predicted by the analytical investigation above.